\documentclass[5p]{elsarticle}
\makeatletter
\def\ps@pprintTitle{%
 \let\@oddhead\@empty
 \let\@evenhead\@empty
 \def\@oddfoot{}%
 \let\@evenfoot\@oddfoot}
\makeatother

\usepackage{csquotes}
\usepackage{graphicx,color}
\usepackage{float}
\usepackage[caption=false]{subfig}
\usepackage{amsmath}
\usepackage{amssymb}
\usepackage{multirow}
\usepackage{dsfont}
\usepackage{adjustbox}
\usepackage{pdflscape}

\begin{document}

\title{Non-retarded room temperature Hamaker constants between elemental metals}
\author{P. Tolias}
\address{Space and Plasma Physics - KTH Royal Institute of Technology, Teknikringen 31, 10044 Stockholm, Sweden}
\begin{abstract}
\noindent The Lifshitz theory of van der Waals forces is utilized for the systematic calculation of the non-retarded room temperature Hamaker constants between $26$ identical isotropic elemental metals that are embedded in vacuum or in pure water. The full spectral method, complemented with a Drude-like low frequency extrapolation, is employed for the elemental metals benefitting from the availability of extended-in-frequency reliable dielectric data. The simple spectral method is employed for pure water and three dielectric representations are explored. Numerical truncation and low frequency extrapolation effects are shown to be negligible. The accuracy of common Lifshitz approximations is quantified. The Hamaker constants for $100$ metal combinations are reported; the geometric mixing rule is revealed to be highly accurate in vacuum \& water.
\end{abstract}
\maketitle

\section{Introduction}\label{intro}

\noindent Surface forces exerted between bodies that lie in close proximity play a pivotal role in most colloid and interface phenomena. In particular, the van der Waals forces and their omnipresent London dispersion force component are important in colloidal stability\,\cite{introd01,introd02}, powder adhesion\,\cite{introd03,introd04}, liquid solid wettability\,\cite{introd05,introd06}, thin liquid film stability and evolution\,\cite{introd07,introd08}, biological interactions\,\cite{introd09}, non-contact friction\,\cite{introd10,introd11} and nanosystem design or manipulation\,\cite{introd12}.

Lifshitz theory allows for a rigorous calculation of van der Waals forces\,\cite{Lifshit1,Lifshit2,Lifshit3,Lifshit4,Lifshit5,Lifchem1,Lifchem2,Lifquan1,Lifquan2,Lifquan3}. In the formalism, the Hamaker coefficient emerges isolating all complexities which originate from cumulative interactions between the instantaneously induced or permanent multipoles that arise inside the bodies and are mediated by the ambient. The Hamaker coefficient, that becomes constant for negligible retardation effects, ultimately depends on relative spectral mismatches in the magneto-dielectric responses between the bodies and the intervening medium.

Despite the fact that the theoretical foundations were laid in the 1940s and the theoretical formalism was crystallized by the 1960s, concrete calculations of Hamaker constants remained sparse. The landscape drastically changed in the last decades due to the development of new instruments or techniques allowing for precise measurements\,\cite{AFMrefe1,AFMrefe2}, the improvement of technical abilities to measure the dielectric spectra necessary for computation\,\cite{optical1,optical2,opticalB}, the advances in ab-initio electronic structure calculations that lead to theoretical dielectric spectra\,\cite{abiniti1,abiniti2,Lifshit6} and miniaturization of technological components to micron or nanoscales. As a consequence, systematic investigations of Hamaker constants are nowadays available targeting ceramic systems\,\cite{introd13,introd14}, ionic compounds\,\cite{introd15}, polymers\,\cite{Metalli1}, biological matter\,\cite{abiniti1} and single wall carbon nanotubes\,\cite{introd16}.

In spite of novel medical applications of metal nanoparticles that concern regimes where the van der Waals forces are dominant\,\cite{introd17,introd18} and the significance of van der Waals interactions in the adhesion between rough metal surfaces \cite{introd19,introd20,introd21}, few theoretical investigations of the Hamaker constants of metals are available in the literature\,\cite{introd22,introd23}. The present contribution aims to fill this void.

In this work, exact Lifshitz calculations are reported for the non-retarded room temperature Hamaker constants between $26$ identical isotropic polycrystalline metals that are embedded in vacuum. The non-retarded Lifshitz formalism is applied without any simplifying theoretical approximation, while both dielectric and magnetic contributions are considered. The full spectral method is employed with input from state-of-the-art extended-in-frequency dielectric data that range from the far infra-red up to the soft X-ray region of the electromagnetic spectrum, roughly $50\,$meV-$10\,$keV\,\cite{opticalB}. The upper numerical cut-offs that are necessarily imposed on the infinite series or the improper integrals are verified to lead to negligible errors. The low frequency extrapolation that is necessary for correct application of the Kramers-Kronig expression follows the Drude model, alternative extrapolation methods have also been explored aiming to demonstrate the robustness of the results. The accuracy of the most common Lifshitz theory approximations is checked. Moreover, non-retarded room temperature Hamaker constants are computed between elemental metals immersed in pure water; the simple spectral method is applied for water and three dielectric representations are explored. Finally, exact Lifshitz calculations are reported for the non-retarded room temperature Hamaker constants of $100$ elemental isotropic metal combinations embedded in vacuum \& water and the accuracy of the widespread geometric combining relation is quantified.

\section{Theoretical aspects}\label{theoretical}

\subsection{The non-retarded dielectric Hamaker constant}\label{LifshitzTheorydielectric}

\noindent Let us consider the interactions between two homogeneous isotropic non-magnetic ($\mu=1$) solid bodies, whose dimensions are large compared to their separation, that are embedded in a homogeneous isotropic non-magnetic medium. Non-retarded Lifshitz theory leads to the Hamaker form $f_{\mathrm{d}}=-A_{132}^{\mathrm{d}}/(6\pi{D}^3)$\cite{Lifshit0} of the van der Waals force per unit area with $D$ the separation and $A_{132}^{\mathrm{d}}$ the Hamaker constant given by the \emph{series-integral expression}\,\cite{Lifshit1,Lifshit2,Lifshit3,Lifshit4,Lifshit5}
\begin{align}
A_{132}^{\mathrm{d}}&=-\frac{3}{2}k_{\mathrm{b}}T{\sum_{n=0}^{\infty}}^{\prime}\int_{0}^{\infty}x\ln{\left\{1-\frac{\left[\epsilon_1(\imath\xi_n)-\epsilon_3(\imath\xi_n)\right]}{\left[\epsilon_1(\imath\xi_n)+\epsilon_3(\imath\xi_n)\right]}\times\right.}\nonumber\\
&\quad{\left.\times\frac{\left[\epsilon_2(\imath\xi_n)-\epsilon_3(\imath\xi_n)\right]}{\left[\epsilon_2(\imath\xi_n)+\epsilon_3(\imath\xi_n)\right]}e^{-x}\right\}}dx\,.\label{HamakerLifshitz1}
\end{align}
In the above; the indices $1,2$ refer to the solid bodies while index $3$ refers to the surrounding medium, $\epsilon_j(\imath\xi_n)$ denotes the dielectric function of imaginary argument evaluated at the bosonic Matsubara frequencies $\xi_n=2\pi{n}k_{\mathrm{b}}T/\hbar$, the prime above the series indicates that the $n=0$ term is considered with half-weight, $T$ is the temperature and $k_{\mathrm{b}}$ the Boltzmann constant. An equivalent \emph{double series expression}, convenient for numerical calculations, is obtained by Taylor expanding the logarithm, interchanging the integral \& summation operators and performing the integration,
\begin{align}
A_{132}^{\mathrm{d}}&=+\frac{3}{2}k_{\mathrm{b}}T{\displaystyle\sum_{n=0}^{\infty}}^{\prime}\displaystyle\sum_{m=1}^{\infty}\frac{1}{m^3}\left\{\frac{\left[\epsilon_1(\imath\xi_n)-\epsilon_3(\imath\xi_n)\right]}{\left[\epsilon_1(\imath\xi_n)+\epsilon_3(\imath\xi_n)\right]}\times\quad\quad\right.\nonumber\\
&\quad\left.\times\frac{\left[\epsilon_2(\imath\xi_n)-\epsilon_3(\imath\xi_n)\right]}{\left[\epsilon_2(\imath\xi_n)+\epsilon_3(\imath\xi_n)\right]}\right\}^m\,.\label{HamakerLifshitz2}
\end{align}
Let us restrict the discussion to conducting bodies that are surrounded by vacuum $\epsilon_3(\omega)\equiv1$. The so-called entropic or static $n=0$ term can be analytically calculated courtesy of $\epsilon_1(0),\epsilon_2(0)\to\infty$ and $\sum_{m=1}^{\infty}(1/m^3)=\zeta(3)$ where $\zeta(.)$ denotes the Riemann zeta function\,\cite{Metalli1,Metalli2}. In this case, the Hamaker constant, $A_{12}^{\mathrm{d}}\equiv{A}_{\mathrm{1v2}}^{\mathrm{d}}$, is given by
\begin{align}
A_{12}^{\mathrm{d}}&=\frac{3}{4}\zeta(3)k_{\mathrm{b}}T+\frac{3}{2}k_{\mathrm{b}}T\sum_{n=1}^{\infty}\sum_{m=1}^{\infty}\frac{1}{m^3}\left\{\frac{\left[\epsilon_1(\imath\xi_n)-1\right]}{\left[\epsilon_1(\imath\xi_n)+1\right]}\times\,\,\right.\nonumber\\
&\quad\left.\times\frac{\left[\epsilon_2(\imath\xi_n)-1\right]}{\left[\epsilon_2(\imath\xi_n)+1\right]}\right\}^m\,.\label{HamakerLifshitz3}
\end{align}
It is instructive to reiterate the major assumptions invoked thus far. \textbf{(1)} Retardation effects due to the finite speed of light have been neglected. This implies that the separation between the bodies is much smaller than the characteristic wavelengths of the relevant absorption spectra\,\cite{Lifshit2}, \emph{i.e.} $D\ll\lambda_0$ or $D\ll(c\hbar)/(\hbar\omega_0)$. \textbf{(2)} Overlapping of electronic wavefunctions has been neglected. This is evident from the use of individual response functions for each body and implies that the separation exceeds few $\mathrm{\AA}$s\,\cite{Lifshit6}, $D\gtrsim0.5$\,nm. \textbf{(3)} Spatial dispersion has been neglected. The long wavelength limit $(k\to0)$ of the dielectric function is implicitly assumed, since only electromagnetic modes with $k\sim1/D$ can provide dominant contributions\,\cite{Lifshit7}. This implies that $D\gtrsim2$\,nm\,\cite{Lifshit8}, otherwise EM modes of wavelength that is comparable to interatomic spacings become important and non-local effects cannot be ignored\,\cite{Lifshit9}. It is worth pointing out that the incorporation of retardation and spatial dispersion ultimately results to a Hamaker coefficient that is not a constant but depends on the separation.

\subsection{The non-retarded magnetic Hamaker constant}\label{LifshitzTheorymagnetic}

\noindent Let us consider the interactions between two homogeneous isotropic magnetic ($\mu\neq1$) solid bodies that are embedded in a homogeneous isotropic magnetic medium. In the non-retarded limit, Lifshitz theory naturally decomposes the Hamaker constant into two independent contributions, \emph{i.e.} $A_{132}=A_{132}^{\mathrm{d}}+A_{132}^{\mathrm{m}}$\,\cite{Lifshit3}. The pure dielectric contribution is acquired by Eq.\ref{HamakerLifshitz2}, whereas the pure magnetic contribution is obtained by the analogous expression\,\cite{Metalli2}
\begin{align}
A_{132}^{\mathrm{m}}&=+\frac{3}{2}k_{\mathrm{b}}T{\displaystyle\sum_{n=0}^{\infty}}^{\prime}\displaystyle\sum_{m=1}^{\infty}\frac{1}{m^3}\left\{\frac{\left[\mu_1(\imath\xi_n)-\mu_3(\imath\xi_n)\right]}{\left[\mu_1(\imath\xi_n)+\mu_3(\imath\xi_n)\right]}\times\quad\quad\right.\nonumber\\
&\quad\left.\times\frac{\left[\mu_2(\imath\xi_n)-\mu_3(\imath\xi_n)\right]}{\left[\mu_2(\imath\xi_n)+\mu_3(\imath\xi_n)\right]}\right\}^m\,.\label{HamakerLifshitz4}
\end{align}
The pure magnetic-induced force per unit area is given by $f_{\mathrm{m}}=-A_{132}^{\mathrm{m}}/(6\pi{D}^3)$\cite{magneti0}. We focus on metal bodies in vacuum $\mu_3(\omega)\equiv1$. The first Matsubara frequency becomes $\xi_{1}\simeq4\times10^{13}\,$Hz at room temperature, while the magnetic permeability is typically characterized by a single relaxation frequency below $f_{\mathrm{rel}}\simeq10^{10}\,$Hz\,\cite{magneti1,magneti2}. As a result of $\xi_1\gg{f}_{\mathrm{rel}}$, $\mu(\imath\xi_n)=1$ can be invoked for $n\geq1$. Therefore, only the static $n=0$ term survives leading to\,\cite{magneti1}
\begin{align}
A_{12}^{\mathrm{m}}&=\frac{3}{4}k_{\mathrm{b}}T\sum_{m=1}^{\infty}\frac{1}{m^3}\left[\frac{\left(\mu_1-1\right)}{\left(\mu_1+1\right)}\frac{\left(\mu_2-1\right)}{\left(\mu_2+1\right)}\right]^m\,.\quad\quad\quad\,\,\,\label{HamakerLifshitz5}
\end{align}
Representative values have been provided in Table \ref{hamakerrecommended_magnetic}. Note that $A_{12}^{\mathrm{d}}\gg{A}_{12}^{\mathrm{m}}$ is valid for elemental metal combinations. In particular, the magnetic contribution is at least nine orders of magnitude lower than the dielectric contribution for paramagnetic or diamagnetic metals, whereas the magnetic contribution is at least two orders of magnitude lower than the dielectric contribution for ferromagnetic metals.

\begin{table}[!t]
\centering
\caption{Magnetic contribution to non-retarded Hamaker constants between $26$ identical elemental polycrystalline metals in vacuum, as computed from Lifshitz theory, see Eq.\ref{HamakerLifshitz5}. In the case of paramagnetic and diamagnetic materials; tabulations of the molar magnetic susceptibility $\chi_{\mathrm{m}}$ were employed\,\cite{magneti3}, the volume magnetic susceptibility was calculated from $\chi_{\mathrm{v}}=(\rho/M)\chi_{\mathrm{m}}$ with $\rho$ the mass density \& $M$ the molar mass and the relative magnetic permeability was then extracted from $\mu=1+4\pi\chi_{\mathrm{v}}$. In the case of ferromagnetic materials; the maximum magnetic permeability of high purity samples was used, \emph{i.e.} $\mu=100000$ for iron, $\mu=600$ for nickel, $\mu=250$ for cobalt.}\label{hamakerrecommended_magnetic}
\begin{tabular}{c c c}
\\ \hline\hline
Metal    & $\chi_{\mathrm{v}}$ (cgs)    & $A_{11}^{\mathrm{m}}$ (Joule)       \\ \hline\hline
Ag		 &  $-1.896\times10^{-6}$       & $4.410\times10^{-31}$	              \\
Al		 &  $+1.651\times10^{-6}$		& $3.343\times10^{-31}$	              \\
Au		 &  $-2.744\times10^{-6}$	    & $9.232\times10^{-31}$               \\
Ba		 &  $+5.265\times10^{-7}$		& $3.400\times10^{-32}$               \\			
Be		 &  $-1.847\times10^{-6}$	    & $4.186\times10^{-31}$               \\
Co		 &  $+1.981\times10^{+1}$	   	& $3.655\times10^{-21}$               \\
Cr		 &  $+2.309\times10^{-5}$		& $6.538\times10^{-29}$               \\
Cu		 &  $-7.699\times10^{-7}$		& $7.269\times10^{-32}$               \\
Fe		 &  $+7.958\times10^{+3}$    	& $3.734\times10^{-21}$               \\				
Hf		 &  $+5.294\times10^{-6}$		& $3.437\times10^{-30}$               \\
Ir		 &  $+2.934\times10^{-6}$		& $1.056\times10^{-30}$               \\
Mo		 &  $+7.714\times10^{-6}$		& $7.297\times10^{-30}$               \\
Nb		 &  $+1.919\times10^{-5}$		& $4.514\times10^{-29}$               \\
Ni		 &  $+4.767\times10^{+1}$		& $3.701\times10^{-21}$               \\	
Os		 &  $+1.306\times10^{-6}$		& $2.093\times10^{-31}$               \\
Pd		 &  $+6.101\times10^{-5}$		& $4.561\times10^{-28}$               \\
Pt		 &  $+2.122\times10^{-5}$	    & $5.521\times10^{-29}$	              \\
Rh		 &  $+1.230\times10^{-5}$	    & $1.855\times10^{-29}$	              \\
Sc		 &  $+1.960\times10^{-5}$		& $4.710\times10^{-29}$               \\
Sr		 &  $+2.772\times10^{-6}$		& $9.423\times10^{-31}$               \\
Ta		 &  $+1.420\times10^{-5}$		& $2.474\times10^{-29}$               \\
Ti		 &  $+1.421\times10^{-5}$ 		& $2.477\times10^{-29}$               \\
Tm		 &  $+4.891\times10^{-6}$		& $2.934\times10^{-30}$               \\		
V 		 &  $+3.418\times10^{-5}$ 	    & $1.432\times10^{-28}$               \\
W 		 &  $+5.564\times10^{-6}$	    & $3.796\times10^{-30}$               \\
Zr	     &  $+8.577\times10^{-6}$ 		& $9.020\times10^{-30}$               \\ \hline\hline
\end{tabular}
\end{table}

\subsection{Approximate non-retarded Hamaker constants}\label{LifshitzTheoryApproximations}

\noindent Aiming to reduce the computational complexity, different approximate forms of the non-retarded dielectric Hamaker constant have been derived in the literature. It is instructive to introduce the most common approximations and to discuss their physical meaning. 

Within the \emph{low temperature approximation}\,\cite{Lifshit2,Lifshit3}, it is implicitly assumed that induced electromagnetic fluctuations are predominantly of quantum rather than classical nature. This is formally correct when the thermal energies are much smaller than the photon energies that correspond to the absorption spectra, \emph{i.e.} $k_{\mathrm{b}}T\ll\hbar\omega_0$. In the $T\to0$ limit, the spacing between successive Matsubara frequencies is infinitesimal $\xi_{n+1}-\xi_{n}=2\pi(k_{\mathrm{b}}T/\hbar)$ and the infinite $n-$series is transformed to an integral. For systems where the static term is important, it is preferable to first isolate the $n=0$ term and then to replace the summation with the corresponding integral\,\cite{Lifshit3}. This procedure leads to
\begin{align}
A_{12}^{\mathrm{d}}&=\frac{3}{4}\zeta(3)k_{\mathrm{b}}T+\frac{3\hbar}{4\pi}\int_{\xi_1}^{\infty}\sum_{m=1}^{\infty}\frac{1}{m^3}\left\{\frac{\left[\epsilon_1(\imath\xi)-1\right]}{\left[\epsilon_1(\imath\xi)+1\right]}\times\,\,\right.\nonumber\\
&\quad\left.\times\frac{\left[\epsilon_2(\imath\xi)-1\right]}{\left[\epsilon_2(\imath\xi)+1\right]}\right\}^md\xi\,.\label{HamakerLifshitz6}
\end{align}
Within the \emph{first order (dipole) approximation}\,\cite{Lifshit5,Metalli1}, only the first term is retained in the Taylor expansion of the logarithm or equivalently only the $m=1$ term is considered in the infinite $m-$series. For systems where the static term is important, it is again preferable to first consider the full $n=0$ term and then to apply the dipole approximation to the remaining terms of the infinite $n-$series. This leads to
\begin{align}
A_{12}^{\mathrm{d}}&=\frac{3}{4}\zeta(3)k_{\mathrm{b}}T+\frac{3}{2}k_{\mathrm{b}}T\sum_{n=1}^{\infty}\left\{\frac{\left[\epsilon_1(\imath\xi_n)-1\right]}{\left[\epsilon_1(\imath\xi_n)+1\right]}\times\right.\,\,\quad\quad\quad\,\,\,\,\nonumber\\
&\quad\times\left.\frac{\left[\epsilon_2(\imath\xi_n)-1\right]}{\left[\epsilon_2(\imath\xi_n)+1\right]}\right\}\,.\label{HamakerLifshitz7}
\end{align}
Within the \emph{low temperature dipole approximation}\,\cite{Metalli1,dielect0}, both limits are simultaneously applied leading to
\begin{align}
A_{12}^{\mathrm{d}}&=\frac{3}{4}\zeta(3)k_{\mathrm{b}}T+\frac{3\hbar}{4\pi}\int_{\xi_1}^{\infty}\frac{\left[\epsilon_1(\imath\xi)-1\right]}{\left[\epsilon_1(\imath\xi)+1\right]}\frac{\left[\epsilon_2(\imath\xi)-1\right]}{\left[\epsilon_2(\imath\xi)+1\right]}d\xi\label{HamakerLifshitz8}
\end{align}

\section{Computational aspects}\label{computational}

\subsection{Dielectric functions of imaginary argument}\label{NumericalD}

\noindent In Lifshitz theory, computation of the non-retarded dielectric Hamaker constant requires evaluation of the imaginary argument dielectric response at the bosonic Matsubara frequencies, \emph{i.e.} $\epsilon(\imath\xi_n)$. Two alternative methods can be employed to determine the $\epsilon(\imath\xi_n)$ Hamaker input from available experimental data of the complex dielectric function.

The most accurate approach is based on the Kramers-Kronig causality relations and is often referred to as \emph{full spectral method}\,\cite{dielect1}. The basic expression reads as\,\cite{LandauB1,LandauB2}
\begin{equation}
\epsilon(\imath\xi_n)=1+\frac{2}{\pi}\int_0^{\infty}\frac{\omega\Im\{\epsilon(\omega)\}}{\omega^2+\xi_n^2}d\omega\,.\label{KramersKronig}
\end{equation}
The method only requires experimental data for the imaginary part of the dielectric function and directly computes $\epsilon(\imath\xi_n)$. Note that combination of Eq.\ref{KramersKronig} with $\omega\Im\{\epsilon(\omega)\}\geq0$ directly proves that $\epsilon(\imath\omega)$ is a monotonically decreasing function asymptotically reaching unity, while elementary Fourier transform manipulations lead to $\epsilon(\imath0)=\epsilon(0)$ that becomes infinite for metals. Hence, $\epsilon(\imath\omega)$ can be labelled as a structure-less function of the frequency, when compared to the non-monotonic $\Im\{\epsilon(\omega)\}$ and $\Re\{\epsilon(\omega)\}$ functions\,\cite{dielect0}. The full spectral method was essentially proposed in the seminal work of Lifshitz\,\cite{Lifshit1} and then mentioned in dedicated early reviews\,\cite{Lifshit2,Lifshit3,dielect2}. Unfortunately, the method relies on a comprehensive spectral characterization of all the involved materials. Consequently, more than a decade intervened between the original suggestion and first application of the method\,\cite{dielect3}. Further systematic applications appeared rather recently but still required extrapolations and interpolations\,\cite{introd13,introd15}.

\begin{table*}[!t]
\centering
\caption{Dominant dielectric contribution to non-retarded Hamaker constants between $26$ identical elemental polycrystalline metals embedded in vacuum. Full spectral method computations without invoking approximations ($A_{\mathrm{full}}^{\mathrm{d}}$, see Eq.\ref{HamakerLifshitz3}), within the low temperature approximation ($A_{\mathrm{lt}}^{\mathrm{d}}$, see Eq.\ref{HamakerLifshitz6}), within the first order approximation ($A_{\mathrm{dp}}^{\mathrm{d}}$, see Eq.\ref{HamakerLifshitz7}) and within the low temperature dipole approximation ($A_{\mathrm{lt,dp}}^{\mathrm{d}}$, see Eq.\ref{HamakerLifshitz8}). Basic characteristics of the dielectric data corresponding to each element are provided in the first columns. Estimates of the residual resulting from the $m-$series truncation of the exact form are provided in the last column.}\label{hamakerrecommended_identical}
\begin{tabular}{c c c c c c c c}
\\ \hline\hline
Metal    & Data range     & Data        & $A_{\mathrm{full}}^{\mathrm{d}}$  & $A_{\mathrm{lt}}^{\mathrm{d}}$ & $A_{\mathrm{dp}}^{\mathrm{d}}$ & $A_{\mathrm{lt,dp}}^{\mathrm{d}}$ & $A_{\mathrm{res}}^{\mathrm{d}}$   \\
         & (eV)           & points      & (Joule)                           & (Joule)                        & (Joule)                        & (Joule)                           & (Joule)                           \\ \hline\hline
Ag       & $0.125-10000$  & 380         & $3.682\times10^{-19}$             & $3.645\times10^{-19}$          & $3.449\times10^{-19}$          & $3.418\times10^{-19}$             & $9.904\times10^{-26}$             \\
Al       & $0.040-10000$  & 254         & $3.554\times10^{-19}$             & $3.516\times10^{-19}$          & $3.220\times10^{-19}$          & $3.189\times10^{-19}$             & $2.031\times10^{-25}$             \\
Au       & $0.125-10000$  & 431         & $4.018\times10^{-19}$             & $3.981\times10^{-19}$          & $3.760\times10^{-19}$          & $3.729\times10^{-19}$             & $8.757\times10^{-26}$             \\
Ba       & $0.300-10000$  & 261         & $1.799\times10^{-19}$             & $1.763\times10^{-19}$          & $1.669\times10^{-19}$          & $1.638\times10^{-19}$             & $1.966\times10^{-26}$             \\
Be       & $0.020-10000$  & 273         & $3.556\times10^{-19}$             & $3.519\times10^{-19}$          & $3.258\times10^{-19}$          & $3.227\times10^{-19}$             & $1.012\times10^{-25}$             \\
Co       & $0.062-10000$  & 275         & $4.109\times10^{-19}$             & $4.072\times10^{-19}$          & $3.829\times10^{-19}$          & $3.798\times10^{-19}$             & $4.914\times10^{-26}$             \\
Cr       & $0.012-10000$  & 299         & $3.709\times10^{-19}$             & $3.672\times10^{-19}$          & $3.431\times10^{-19}$          & $3.400\times10^{-19}$             & $7.945\times10^{-26}$             \\
Cu       & $0.130-10000$  & 343         & $3.382\times10^{-19}$             & $3.344\times10^{-19}$          & $3.157\times10^{-19}$          & $3.126\times10^{-19}$             & $7.818\times10^{-26}$             \\
Fe       & $0.022-10000$  & 419         & $3.883\times10^{-19}$             & $3.846\times10^{-19}$          & $3.608\times10^{-19}$          & $3.577\times10^{-19}$             & $6.853\times10^{-26}$             \\
Hf       & $0.300-10000$  & 267         & $2.755\times10^{-19}$             & $2.721\times10^{-19}$          & $2.605\times10^{-19}$          & $2.575\times10^{-19}$             & $1.752\times10^{-27}$             \\
Ir       & $0.034-10000$  & 304         & $5.342\times10^{-19}$             & $5.304\times10^{-19}$          & $4.940\times10^{-19}$          & $4.909\times10^{-19}$             & $1.358\times10^{-25}$             \\
Mo       & $0.010-10000$  & 434         & $4.806\times10^{-19}$             & $4.769\times10^{-19}$          & $4.431\times10^{-19}$          & $4.400\times10^{-19}$             & $1.318\times10^{-25}$             \\
Nb       & $0.120-10000$  & 407         & $4.638\times10^{-19}$             & $4.601\times10^{-19}$          & $4.299\times10^{-19}$          & $4.268\times10^{-19}$             & $1.013\times10^{-25}$             \\
Ni       & $0.100-10000$  & 352         & $3.692\times10^{-19}$             & $3.655\times10^{-19}$          & $3.447\times10^{-19}$          & $3.416\times10^{-19}$             & $4.997\times10^{-26}$             \\
Os       & $0.100-10000$  & 399         & $4.796\times10^{-19}$             & $4.760\times10^{-19}$          & $4.477\times10^{-19}$          & $4.446\times10^{-19}$             & $2.470\times10^{-26}$             \\
Pd       & $0.100-10000$  & 319         & $3.886\times10^{-19}$             & $3.850\times10^{-19}$          & $3.641\times10^{-19}$          & $3.610\times10^{-19}$             & $3.203\times10^{-26}$             \\
Pt       & $0.100-10000$  & 368         & $4.501\times10^{-19}$             & $4.464\times10^{-19}$          & $4.196\times10^{-19}$          & $4.165\times10^{-19}$             & $6.684\times10^{-26}$             \\
Rh       & $0.100-10000$  & 336         & $4.559\times10^{-19}$             & $4.521\times10^{-19}$          & $4.223\times10^{-19}$          & $4.192\times10^{-19}$             & $1.146\times10^{-25}$             \\
Sc       & $0.270-10000$  & 275         & $2.340\times10^{-19}$             & $2.304\times10^{-19}$          & $2.181\times10^{-19}$          & $2.151\times10^{-19}$             & $1.005\times10^{-26}$             \\
Sr       & $0.300-10000$  & 259         & $2.194\times10^{-19}$             & $2.158\times10^{-19}$          & $2.042\times10^{-19}$          & $2.012\times10^{-19}$             & $2.978\times10^{-26}$             \\
Ta       & $0.010-10000$  & 430         & $4.495\times10^{-19}$             & $4.458\times10^{-19}$          & $4.174\times10^{-19}$          & $4.143\times10^{-19}$             & $1.069\times10^{-25}$             \\
Ti       & $0.0062-10000$ & 293         & $2.668\times10^{-19}$             & $2.632\times10^{-19}$          & $2.493\times10^{-19}$          & $2.463\times10^{-19}$             & $2.934\times10^{-26}$             \\
Tm       & $0.062-10000$  & 266         & $3.311\times10^{-19}$             & $3.274\times10^{-19}$          & $3.070\times10^{-19}$          & $3.039\times10^{-19}$             & $3.815\times10^{-26}$             \\
V        & $0.138-10000$  & 296         & $3.536\times10^{-19}$             & $3.500\times10^{-19}$          & $3.288\times10^{-19}$          & $3.257\times10^{-19}$             & $3.467\times10^{-26}$             \\
W        & $0.140-10000$  & 479         & $5.106\times10^{-19}$             & $5.069\times10^{-19}$          & $4.741\times10^{-19}$          & $4.710\times10^{-19}$             & $7.747\times10^{-26}$             \\
Zr       & $0.073-10000$  & 495         & $3.094\times10^{-19}$             & $3.057\times10^{-19}$          & $2.893\times10^{-19}$          & $2.862\times10^{-19}$             & $3.386\times10^{-26}$             \\ \hline\hline
\end{tabular}
\end{table*}

On the other hand, the \emph{simple spectral method} is based on analytical parameterizations of the dielectric function within a restricted spectral range\,\cite{dielect1}. To be more specific, a model dielectric function is assumed that typically combines a Debye relaxation term with a sum of Lorentz oscillators for dielectric media\,\cite{paramet1,paramet2} or a Drude free electron term with a sum of Lorentz or Brendel-Bormann oscillators for conducting media\,\cite{paramet3,paramet4}. The unknown resonant frequencies, oscillator strengths and damping constants are then determined by simultaneous fits to experimental data for the real and imaginary parts of the dielectric function. Finally, $\epsilon(\imath\xi_n)$ is obtained by a direct $\omega\to\imath\xi_n$ substitution.

The main disadvantage of the full spectral method lies in the requirement of extended-in-frequency dielectric data for the accurate evaluation of $\epsilon(\imath\xi_n)$, as a result of the infinite upper integration limit present in the Kramers-Kronig relation. On the other hand, the main disadvantage of the simple spectral method lies in the difficulty of curve-fitting dielectric data that feature multiple structural features in extended frequency ranges. Thus, it is obvious that the full spectral method is suitable for precise determinations of the non-retarded Hamaker constants of well-characterized materials, whereas the simple spectral method is suitable for rough estimates of the non-retarded Hamaker constants of poorly-characterized materials.

For the polycrystalline metals of interest, experimental room temperature dielectric data are available from the far infra-red up to the soft X-ray region of the electromagnetic spectrum, roughly $50\,$meV-$10\,$keV\,\cite{opticalB}. The large amount and extended range of the available dielectric data suffice for application of the full spectral method.

\subsection{Numerical input}\label{NumericalI}

\noindent The necessary experimental room temperature ($300\,$K) dielectric data are adopted from Adachi's handbook\,\cite{opticalB} that contains extended tabulations of the long wavelength relative permittivity of $63$ elemental metals as function of the frequency. The number of relevant datasets was reduced to $52$, when focusing only on isotropic polycrystalline solids. In addition, $26$ datasets were considered to be inappropriate for accurate Lifshitz calculations. In particular, these datasets were excluded from analysis due to the presence of rather extended frequency gaps ($16$ elements), the lack of low frequency measurements near the infrared range ($6$ elements) and the rather sparse visible or ultraviolet measurements ($4$ elements). Overall, extended room temperature dielectric data were available for $26$ elemental metals. It is worth pointing out that, in spite of the fact that the imaginary part of the dielectric function $\Im\{\epsilon\}$ constituted one of the tabulated quantities\,\cite{opticalB}, it was also calculated with the aid of tabulated complex refractive index $n+\imath\kappa$ data through the basic relation $\Im\{\epsilon\}=2n\kappa$ in an effort to detect the occurrence of misprints.

Naturally, the list of $26$ metals includes the most common and most technologically important elemental metals. The highest EM frequency that is available corresponds to $\hbar\omega=10000\,$eV for all metals, while the lowest frequency available varies from $\hbar\omega=0.0062\,$eV up to $\hbar\omega=0.300\,$eV. The number of data points ranges from $254$ up to $495$. The frequency intervals and data points corresponding to each element can be found in Table \ref{hamakerrecommended_identical}.

\subsection{Numerical approximations}\label{NumericalA}

\noindent Lifshitz theory calculations of the dielectric contribution to the non-retarded Hamaker constant performed with the double series expression and the full spectral method, Eq.\ref{HamakerLifshitz3} and Eq.\ref{KramersKronig}, involve the computation of an improper integral and two infinite series. The mathematical operations need to be truncated in a manner that ensures minimum errors.

The \emph{improper integral} of the Kramers-Kronig type expression requires knowledge of the imaginary part of the dielectric function at all frequencies. The available dielectric data extend to high enough frequencies so that upper extrapolations are not necessary. Here, an upper integration limit of $\hbar\omega_{\mathrm{u}}=10000\,$eV has been imposed and it has been verified that upper limits at least down to $\hbar\omega_{\mathrm{u}}=5000\,$eV lead to indistinguishable results. On the other hand, owing to $\epsilon(0)\to\infty$ for metals, Hamaker constants are sensitive to the lower integration limit and lower extrapolations down to zero frequencies are necessary. The extrapolation procedure is sketched in what follows: \textbf{(a)} Bound-electron inter-band effects are assumed to be completely damped roughly below $0.6\,$eV and only free-electron intra-band effects are assumed to dominate the low frequency dielectric response. These effects are parameterized with the aid of the Drude model, \emph{i.e.} $\epsilon_{\mathrm{D}}(\omega)=1-(f\omega_{\mathrm{p}}^2)/[\omega(\omega+\imath\Gamma)]$ with $f$ a coupling strength, $\omega_{\mathrm{p}}$ the plasma frequency and $\Gamma$ the damping constant. This directly results to $\Im\{\epsilon_{\mathrm{D}}(\omega)\}=a/[\omega(\omega^2+b^2)]$ after setting $a=f\omega_{\mathrm{p}}^2\Gamma$, $b=\Gamma$. \textbf{(b)} The unknown parameters $a,b$ are determined by least-square fitting to the low frequency data $\hbar\omega\leq0.6$eV. The number of relevant data points varied from $4$ to $69$ depending on the element. The Drude model is employed only in the extrapolated range and not also in the fitting range. \textbf{(c)} After some algebraic manipulations and with the aid of contour integration, the Eq.\ref{KramersKronig} Kramers-Kronig expression is ultimately rewritten as
\begin{equation}
\epsilon(\imath\xi_n)=\epsilon_{\mathrm{D}}(\imath\xi_n)+\frac{2}{\pi}\int_{\omega_{\mathrm{l}}}^{\omega_{\mathrm{u}}}\frac{\omega\left[\Im\{\epsilon(\omega)-\epsilon_{\mathrm{D}}(\omega)\}\right]}{\omega^2+\xi_n^2}d\omega\,,\label{KramersKronigNew}
\end{equation}
where $\omega_{\mathrm{u}},\,\omega_{\mathrm{l}}$ are the highest and the lowest frequencies in the dielectric data of each element, respectively. Hermite polynomial interpolation schemes are then utilized under the physical constraint $\Im\{\epsilon(\omega)\}>0$ in order to construct an analytic $\Im\{\epsilon(\omega)\}$ representation. Finally, the integration in Eq.\ref{KramersKronigNew} is carried out numerically with the Gauss-Kronrod rule. Alternative quadrature methods have also been utilized, confirming the accuracy of the final result.

The primary \emph{n-series} allows for computation of the additive contributions to the non-retarded Hamaker constant that stem from different electromagnetic frequency ($\omega$) or equivalently photon energy ($\hbar\omega$) ranges. As a consequence of the monotonic $\epsilon(\imath\omega)$ decrease with increasing frequency, contributions from increasing Matsubara frequencies gradually decrease. Within the neighborhood of $\hbar\xi_n\simeq300\,$eV for most metals, $\epsilon(\imath\xi_n)$ has nearly reached its asymptotic limit of unity and higher frequency contributions become negligible. In the present calculations, aiming to take utmost advantage of the available dielectric data, the n-series is truncated at $n=61564$ which corresponds to the last Matsubara frequency prior to $10000\,$eV. The residual contributions from all neglected terms are expected to be at least six orders of magnitude lower.

The secondary \emph{m-series} should converge quite fast considering the satisfactory accuracy of the first-order approximation. Convergence rates should increase with the Matsubara frequency given the $\epsilon(\imath\xi_n)$ decrease towards unity. In the present calculations, the m-series is truncated at $m=50$ regardless of the Matsubara frequency. The residual contributions from all neglected terms are expected to be at least six orders of magnitude lower.

Aiming at an independent confirmation of the accuracy, Lifshitz theory calculations were also carried out with the series-integral expression and the full spectral method, see Eq.\ref{HamakerLifshitz1} and Eq.\ref{KramersKronig}. This equivalent procedure involves the computation of two improper integrals and one infinite series. Following expectations, the results turned out to be identical up to five significant figures. Note that the computational cost of the series-integral expression turned out to be comparable with that of the double series expression.

\section{Results}\label{results}

\subsection{Identical metals in vacuum}\label{identical}

\noindent The non-retarded room temperature dielectric Hamaker constants of van der Waals interactions between identical isotropic elemental metals separated by vacuum are listed in Table \ref{hamakerrecommended_identical}. The full spectral method results employing the Drude low frequency extrapolation without invoking any further approximations vary between $1.799\times10^{-19}\,$J (barium) and $5.342\times10^{-19}\,$J (iridium) for the $26$ elements investigated. The low temperature approximation of Eq.\ref{HamakerLifshitz6} is revealed to be very accurate, only exhibiting $1.05\%$, $2.00\%$, $0.71\%$ for the mean, maximum and minimum relative deviations compared to the exact result. The first-order approximation of Eq.\ref{HamakerLifshitz7} is deduced to be noticeably less accurate, since it exhibits $7.04\%$, $9.40\%$, $5.44\%$ for the mean, maximum and minimum relative deviations with respect to the exact result. The low temperature dipole approximation of Eq.\ref{HamakerLifshitz8} is even less accurate but is still deemed satisfactory, exhibiting $7.92\%$, $10.27\%$, $6.53\%$ for the mean, maximum and minimum relative deviations. Finally, it is worth pointing out that the highest Hamaker constants belong to $4$d and $5$d refractory metals; namely Nb,\,Mo,\,Rh and Ta,\,W,\,Os,\,Ir.

\begin{landscape}
\begin{table}
\centering
\caption{Dominant dielectric contribution to non-retarded Hamaker constants between $26$ identical elemental polycrystalline metals that are embedded in vacuum. Full spectral method computations without invoking any approximations (see Eq.\ref{HamakerLifshitz3}). Results are decomposed into the static term and eight arbitrary frequency intervals. The number of bosonic Matsubara frequencies contained in each interval are the following: $n=30$ in $0-5\,$eV, $n=31$ in $5-10\,$eV, $n=92$ in $10-25\,$eV, $n=154$ in $25-50\,$eV, $n=308$ in $50-100\,$eV, $n=1231$ in $100-300\,$eV, $n=4310$ in $300-1000\,$eV, $n=55408$ in $1000-10000\,$eV.}\label{hamakerrecommended_interval}
\begin{tabular}{c c c c c c c c c c c}
\\ \hline\hline
   & static                & $0-5\,$eV             & $5-10\,$eV            & $10-25\,$eV           & $25-50\,$eV           & $50-100\,$eV          & $100-300\,$eV         & $300-1000\,$eV        & $1000-10000\,$eV      \\ \hline\hline
Ag & $3.734\times10^{-21}$ & $1.535\times10^{-19}$ & $6.771\times10^{-20}$ & $8.105\times10^{-20}$ & $3.957\times10^{-20}$ & $1.689\times10^{-20}$ & $5.325\times10^{-21}$ & $4.421\times10^{-22}$ & $2.577\times10^{-23}$ \\
Al & $3.734\times10^{-21}$ & $1.869\times10^{-19}$ & $9.248\times10^{-20}$ & $5.931\times10^{-20}$ & $9.788\times10^{-21}$ & $2.340\times10^{-21}$ & $7.274\times10^{-22}$ & $6.507\times10^{-23}$ & $2.732\times10^{-24}$ \\
Au & $3.734\times10^{-21}$ & $1.577\times10^{-19}$ & $7.958\times10^{-20}$ & $9.590\times10^{-20}$ & $4.228\times10^{-20}$ & $1.651\times10^{-20}$ & $5.510\times10^{-21}$ & $6.200\times10^{-22}$ & $4.357\times10^{-23}$ \\
Ba & $3.734\times10^{-21}$ & $1.211\times10^{-19}$ & $3.051\times10^{-20}$ & $1.873\times10^{-20}$ & $4.340\times10^{-21}$ & $1.196\times10^{-21}$ & $3.459\times10^{-22}$ & $2.890\times10^{-23}$ & $1.719\times10^{-24}$ \\
Be & $3.734\times10^{-21}$ & $1.738\times10^{-19}$ & $9.018\times10^{-20}$ & $7.072\times10^{-20}$ & $1.369\times10^{-20}$ & $2.824\times10^{-21}$ & $6.246\times10^{-22}$ & $4.020\times10^{-23}$ & $1.385\times10^{-24}$ \\
Co & $3.734\times10^{-21}$ & $1.629\times10^{-19}$ & $8.850\times10^{-20}$ & $9.932\times10^{-20}$ & $3.768\times10^{-20}$ & $1.386\times10^{-20}$ & $4.546\times10^{-21}$ & $3.821\times10^{-22}$ & $1.911\times10^{-23}$ \\
Cr & $3.734\times10^{-21}$ & $1.678\times10^{-19}$ & $8.385\times10^{-20}$ & $7.914\times10^{-20}$ & $2.527\times10^{-20}$ & $8.351\times10^{-21}$ & $2.511\times10^{-21}$ & $2.105\times10^{-22}$ & $1.143\times10^{-23}$ \\
Cu & $3.734\times10^{-21}$ & $1.517\times10^{-19}$ & $6.868\times10^{-20}$ & $7.093\times10^{-20}$ & $2.749\times10^{-20}$ & $1.109\times10^{-20}$ & $4.136\times10^{-21}$ & $3.969\times10^{-22}$ & $2.032\times10^{-23}$ \\
Fe & $3.734\times10^{-21}$ & $1.643\times10^{-19}$ & $8.510\times10^{-20}$ & $8.879\times10^{-20}$ & $3.142\times10^{-20}$ & $1.105\times10^{-20}$ & $3.530\times10^{-21}$ & $3.004\times10^{-22}$ & $1.554\times10^{-23}$ \\
Hf & $3.734\times10^{-21}$ & $1.259\times10^{-19}$ & $5.279\times10^{-20}$ & $5.598\times10^{-20}$ & $2.362\times10^{-20}$ & $9.623\times10^{-21}$ & $3.491\times10^{-21}$ & $3.811\times10^{-22}$ & $2.412\times10^{-23}$ \\
Ir & $3.734\times10^{-21}$ & $1.855\times10^{-19}$ & $1.149\times10^{-19}$ & $1.408\times10^{-19}$ & $5.865\times10^{-20}$ & $2.208\times10^{-20}$ & $7.602\times10^{-21}$ & $8.806\times10^{-22}$ & $6.128\times10^{-23}$ \\
Mo & $3.734\times10^{-21}$ & $1.825\times10^{-19}$ & $1.105\times10^{-19}$ & $1.243\times10^{-19}$ & $4.301\times10^{-20}$ & $1.290\times10^{-20}$ & $3.344\times10^{-21}$ & $3.131\times10^{-22}$ & $1.991\times10^{-23}$ \\
Nb & $3.734\times10^{-21}$ & $1.744\times10^{-19}$ & $1.025\times10^{-19}$ & $1.198\times10^{-19}$ & $4.505\times10^{-20}$ & $1.432\times10^{-20}$ & $3.674\times10^{-21}$ & $3.040\times10^{-22}$ & $1.694\times10^{-23}$ \\
Ni & $3.734\times10^{-21}$ & $1.571\times10^{-19}$ & $7.653\times10^{-20}$ & $8.041\times10^{-20}$ & $3.234\times10^{-20}$ & $1.357\times10^{-20}$ & $5.058\times10^{-21}$ & $4.633\times10^{-22}$ & $2.313\times10^{-23}$ \\
Os & $3.734\times10^{-21}$ & $1.659\times10^{-19}$ & $1.024\times10^{-19}$ & $1.267\times10^{-19}$ & $5.282\times10^{-20}$ & $2.013\times10^{-20}$ & $7.138\times10^{-21}$ & $8.398\times10^{-22}$ & $5.831\times10^{-23}$ \\
Pd & $3.734\times10^{-21}$ & $1.547\times10^{-19}$ & $7.745\times10^{-20}$ & $9.088\times10^{-20}$ & $4.061\times10^{-20}$ & $1.597\times10^{-20}$ & $4.839\times10^{-21}$ & $4.287\times10^{-22}$ & $2.715\times10^{-23}$ \\
Pt & $3.734\times10^{-21}$ & $1.679\times10^{-19}$ & $9.322\times10^{-20}$ & $1.112\times10^{-19}$ & $4.749\times10^{-20}$ & $1.892\times10^{-20}$ & $6.767\times10^{-21}$ & $7.866\times10^{-22}$ & $5.536\times10^{-23}$ \\
Rh & $3.734\times10^{-21}$ & $1.775\times10^{-19}$ & $9.790\times10^{-20}$ & $1.100\times10^{-19}$ & $4.481\times10^{-20}$ & $1.652\times10^{-20}$ & $4.889\times10^{-21}$ & $4.483\times10^{-22}$ & $2.879\times10^{-23}$ \\
Sc & $3.734\times10^{-21}$ & $1.330\times10^{-19}$ & $4.858\times10^{-20}$ & $3.547\times10^{-20}$ & $9.581\times10^{-21}$ & $2.853\times10^{-21}$ & $7.206\times10^{-22}$ & $5.439\times10^{-23}$ & $2.822\times10^{-24}$ \\
Sr & $3.734\times10^{-21}$ & $1.297\times10^{-19}$ & $4.329\times10^{-20}$ & $3.188\times10^{-20}$ & $8.182\times10^{-21}$ & $2.085\times10^{-21}$ & $4.872\times10^{-22}$ & $3.908\times10^{-23}$ & $1.962\times10^{-24}$ \\
Ta & $3.734\times10^{-21}$ & $1.720\times10^{-19}$ & $9.767\times10^{-20}$ & $1.107\times10^{-19}$ & $4.277\times10^{-20}$ & $1.627\times10^{-20}$ & $5.703\times10^{-21}$ & $6.142\times10^{-22}$ & $3.876\times10^{-23}$ \\
Ti & $3.734\times10^{-21}$ & $1.376\times10^{-19}$ & $5.424\times10^{-20}$ & $4.797\times10^{-20}$ & $1.623\times10^{-20}$ & $5.456\times10^{-21}$ & $1.467\times10^{-21}$ & $1.086\times10^{-22}$ & $5.512\times10^{-24}$ \\
Tm & $3.734\times10^{-21}$ & $1.583\times10^{-19}$ & $7.617\times10^{-20}$ & $6.584\times10^{-20}$ & $1.881\times10^{-20}$ & $6.045\times10^{-21}$ & $1.954\times10^{-21}$ & $1.993\times10^{-22}$ & $1.240\times10^{-23}$ \\
V  & $3.734\times10^{-21}$ & $1.585\times10^{-19}$ & $7.953\times10^{-20}$ & $7.542\times10^{-20}$ & $2.500\times10^{-20}$ & $8.704\times10^{-21}$ & $2.578\times10^{-21}$ & $1.999\times10^{-22}$ & $1.006\times10^{-23}$ \\
W  & $3.734\times10^{-21}$ & $1.752\times10^{-19}$ & $1.110\times10^{-19}$ & $1.377\times10^{-19}$ & $5.504\times10^{-20}$ & $2.021\times10^{-20}$ & $6.886\times10^{-21}$ & $7.579\times10^{-22}$ & $4.925\times10^{-23}$ \\
Zr & $3.734\times10^{-21}$ & $1.448\times10^{-19}$ & $6.344\times10^{-20}$ & $6.503\times10^{-20}$ & $2.323\times10^{-20}$ & $7.119\times10^{-21}$ & $1.838\times10^{-21}$ & $1.664\times10^{-22}$ & $9.646\times10^{-24}$ \\ \hline\hline
\end{tabular}
\end{table}
\end{landscape}

Contributions to the Hamaker constant stemming from different frequency ranges of the electromagnetic spectrum are detailed in Table \ref{hamakerrecommended_interval}. Major contributions $\gtrsim90\%$ originate from the infrared and ultraviolet regions of the EM spectrum. The contributions from the photon energy interval of $100-300\,$eV are two orders of magnitude lower than the total and of the same order with the static term, which suggests that extended dielectric data up to the neighborhood of $300\,$eV are required for accurate calculations. The contributions from the photon energy interval of $1000-10000\,$eV are four to five orders of magnitude lower than the total, which justifies ignoring all the Matsubara frequencies beyond that range. In other words, the numerical cut-off of the primary n-series leads to negligible errors. In addition, the contributions to the Hamaker constant that stem from the neglected $m=51$ term have been computed to be six to eight orders of magnitude lower than the total, see the last column of Table  \ref{hamakerrecommended_identical}. It is evident that the numerical cut-off of the secondary m-series also leads to negligible errors.

\begin{table}[!t]
\centering
\caption{Effect of the low frequency extrapolation of the Kramers-Kronig expression (Eq.\ref{KramersKronig}) on the dielectric contribution to the non-retarded Hamaker constants between $26$ identical elemental polycrystalline metals that are embedded in vacuum. A physical extrapolation based on the Drude model is compared to a pure mathematical extrapolation based on polynomials of variable degree.}\label{hamakerrecommended_extrapolation}
\begin{tabular}{c c c c}
\\ \hline\hline
Metal    & $A_{\mathrm{Dru}}^{\mathrm{d}}$  & $A_{\mathrm{poly}}^{\mathrm{d}}$  & Dev        \\
         & (Joule)                         & (Joule)                            & $(\%)$     \\ \hline\hline
Ag       & $3.682\times10^{-19}$           & $3.431\times10^{-19}$              & $+6.82$    \\
Al       & $3.554\times10^{-19}$           & $3.582\times10^{-19}$              & $-0.79$    \\
Au       & $4.018\times10^{-19}$           & $3.923\times10^{-19}$              & $+2.36$    \\
Ba       & $1.799\times10^{-19}$           & $1.761\times10^{-19}$              & $+2.11$    \\
Be       & $3.556\times10^{-19}$           & $3.528\times10^{-19}$              & $+0.79$    \\
Co       & $4.109\times10^{-19}$           & $4.100\times10^{-19}$              & $+0.22$    \\
Cr       & $3.709\times10^{-19}$           & $3.692\times10^{-19}$              & $+0.46$    \\
Cu       & $3.382\times10^{-19}$           & $3.126\times10^{-19}$              & $+7.57$    \\
Fe       & $3.883\times10^{-19}$           & $3.830\times10^{-19}$              & $+1.36$    \\
Hf       & $2.755\times10^{-19}$           & $2.750\times10^{-19}$              & $+0.18$    \\
Ir       & $5.342\times10^{-19}$           & $5.327\times10^{-19}$              & $+0.28$    \\
Mo       & $4.806\times10^{-19}$           & $4.617\times10^{-19}$              & $+3.93$    \\
Nb       & $4.638\times10^{-19}$           & $4.526\times10^{-19}$              & $+2.41$    \\
Ni       & $3.692\times10^{-19}$           & $3.711\times10^{-19}$              & $-0.51$    \\
Os       & $4.796\times10^{-19}$           & $4.792\times10^{-19}$              & $+0.08$    \\
Pd       & $3.886\times10^{-19}$           & $3.898\times10^{-19}$              & $-0.31$    \\
Pt       & $4.501\times10^{-19}$           & $4.458\times10^{-19}$              & $+0.96$    \\
Rh       & $4.559\times10^{-19}$           & $4.506\times10^{-19}$              & $+1.16$    \\
Sc       & $2.340\times10^{-19}$           & $2.376\times10^{-19}$              & $-1.54$    \\
Sr       & $2.194\times10^{-19}$           & $2.156\times10^{-19}$              & $+1.73$    \\
Ta       & $4.495\times10^{-19}$           & $4.477\times10^{-19}$              & $+0.40$    \\
Ti       & $2.668\times10^{-19}$           & $2.663\times10^{-19}$              & $+0.19$    \\
Tm       & $3.311\times10^{-19}$           & $3.248\times10^{-19}$              & $+1.90$    \\
V        & $3.536\times10^{-19}$           & $3.533\times10^{-19}$              & $+0.08$    \\
W        & $5.106\times10^{-19}$           & $4.999\times10^{-19}$              & $+2.10$    \\
Zr       & $3.094\times10^{-19}$           & $3.270\times10^{-19}$              & $-5.69$    \\ \hline\hline
\end{tabular}
\end{table}

Strictly speaking, the non-retarded formalism cannot be applied for the highest frequencies present in the Adachi dataset, especially in the soft X-ray spectral region. The condition for negligible retardation effects $D\ll(c\hbar)/(\hbar\omega_0)$ together with the condition for negligible metallic bonding effects $D\gtrsim0.5\,$nm directly lead to a $\hbar\omega_0\ll400\,$eV applicability condition for non-retarded Lifshitz theory. However, since Hamaker contributions from all the Matsubara frequencies beyond $300\,$eV are insignificant, high frequency retardation effects should be negligible. Hence, it was preferred to retain the non-retarded formalism for all frequencies up to $10000\,$eV.

Finally, the effect of the low frequency extrapolation of the Kramers-Kronig expression is investigated in Table \ref{hamakerrecommended_extrapolation}, where non-retarded Hamaker constants resulting from the physical Drude model and a mathematical polynomial model are compared. Such elementary polynomial models have been used in the past for the computation of Hamaker constants, see for instance the work of Osborne-Lee where the $\Im\{\epsilon(\omega)\}$ of elemental metals was linearly extrapolated down to zero frequencies\,\cite{introd22} or the recent studies of the present author where the $\Im\{\epsilon(\omega)\}$ of elemental metals was quadratically extrapolated down to zero frequencies\,\cite{introd23}. The mean, maximum and minimum absolute relative deviations between the two extrapolation methods are $1.77\%$, $7.57\%$ and $0.08\%$, respectively. Notice that, depending on the element, the polynomial model can lead to an overestimation or underestimation compared to the Drude model. These deviations are rather small but they surpass the errors caused by the low temperature approximation, which highlights the need for a proper physical extrapolation below the far infrared range.

\subsection{Metal combinations in vacuum}\label{different}

\noindent Combining relations or mixing rules are often employed in order to obtain approximate values for unknown Hamaker constants between different combinations of materials in terms of known Hamaker constants between identical materials\,\cite{Metalli1,dielect1,AFMrefe3}. Within the assumption of a vacuum intervening medium, such relations allow for the estimation of the Hamaker constants between $N(N-1)/2$ material combinations given knowledge of the Hamaker constants between $N$ identical materials. The most widespread combining relation is based on the expression for the geometric mean and simply reads as
\begin{equation}
A_{12}^{\mathrm{d}}\simeq\sqrt{A_{11}^{\mathrm{d}}A_{22}^{\mathrm{d}}}\,.\label{combiningrule1}
\end{equation}
The accuracy of the geometric mixing rule has been tested against full Lifshitz computations for $100$ metal combinations, see Table \ref{hamakerrecommended_combination} for details. The relation is revealed to be very accurate, only exhibiting $0.94\%$, $4.58\%$, $0.02\%$ for the mean, maximum and minimum relative deviations with respect to the exact result. By applying the Cauchy-Schwarz inequality to Eq.\ref{HamakerLifshitz3}, it can be proven that geometrical mixing always leads to overestimations.

\begin{table*}
\centering
\caption{Dominant dielectric contribution to the non-retarded Hamaker constants of $100$ elemental polycrystalline metal combinations embedded in vacuum. Full spectral method computations that employ the Drude low frequency extrapolation without invoking any approximations. These first
principle results are compared with the results of the standard combining relation, see Eq.\ref{combiningrule1}. Note that reciprocity applies as also dictated from Newton's third law, $A_{12}=A_{21}$.}\label{hamakerrecommended_combination}
\begin{tabular}{c c c c || c c c c}
\\ \hline\hline
Pair          & $A_{12}^{\mathrm{d}}$ (Joule)   & $\sqrt{A_{11}^{\mathrm{d}}A_{22}^{\mathrm{d}}}$ (Joule) & Dev $(\%)$ & Pair          & $A_{12}^{\mathrm{d}}$ (Joule)       & $\sqrt{A_{11}^{\mathrm{d}}A_{22}^{\mathrm{d}}}$ (Joule) & Dev $(\%)$    \\ \hline\hline
Ag-Al         & $3.478\times10^{-19}$           & $3.617\times10^{-19}$                                   & 4.00       & Fe-Ni         & $3.783\times10^{-19}$               & $3.786\times10^{-19}$                                   & 0.08     \\
Ag-Au         & $3.843\times10^{-19}$           & $3.846\times10^{-19}$                                   & 0.08       & Fe-Ta         & $4.171\times10^{-19}$               & $4.178\times10^{-19}$                                   & 0.17     \\
Ag-Co         & $3.879\times10^{-19}$           & $3.890\times10^{-19}$                                   & 0.28       & Fe-V          & $3.703\times10^{-19}$               & $3.705\times10^{-19}$                                   & 0.05     \\
Ag-Cu         & $3.520\times10^{-19}$           & $3.529\times10^{-19}$                                   & 0.26       & Fe-W          & $4.428\times10^{-19}$               & $4.453\times10^{-19}$                                   & 0.56     \\
Ag-Fe         & $3.766\times10^{-19}$           & $3.781\times10^{-19}$                                   & 0.40       & Hf-Ir         & $3.788\times10^{-19}$               & $3.836\times10^{-19}$                                   & 1.27     \\
Ag-Ir         & $4.408\times10^{-19}$           & $4.435\times10^{-19}$                                   & 0.61       & Hf-Mo         & $3.602\times10^{-19}$               & $3.639\times10^{-19}$                                   & 1.03     \\
Ag-Mo         & $4.176\times10^{-19}$           & $4.207\times10^{-19}$                                   & 0.74       & Hf-Nb         & $3.544\times10^{-19}$               & $3.575\times10^{-19}$                                   & 0.87     \\
Ag-Ni         & $3.682\times10^{-19}$           & $3.687\times10^{-19}$                                   & 0.14       & Hf-Os         & $3.596\times10^{-19}$               & $3.635\times10^{-19}$                                   & 1.08     \\
Ag-Pd         & $3.780\times10^{-19}$           & $3.783\times10^{-19}$                                   & 0.08       & Hf-Ta         & $3.497\times10^{-19}$               & $3.519\times10^{-19}$                                   & 0.63     \\
Ag-Pt         & $4.061\times10^{-19}$           & $4.071\times10^{-19}$                                   & 0.25       & Hf-W          & $3.706\times10^{-19}$               & $3.751\times10^{-19}$                                   & 1.21     \\
Ag-Ta         & $4.055\times10^{-19}$           & $4.068\times10^{-19}$                                   & 0.32       & Ir-Pd         & $4.542\times10^{-19}$               & $4.556\times10^{-19}$                                   & 0.31     \\
Ag-Ti         & $3.097\times10^{-19}$           & $3.134\times10^{-19}$                                   & 1.19       & Ir-Rh         & $4.927\times10^{-19}$               & $4.935\times10^{-19}$                                   & 0.16     \\
Ag-W          & $4.310\times10^{-19}$           & $4.336\times10^{-19}$                                   & 0.60       & Ir-Ti         & $3.682\times10^{-19}$               & $3.775\times10^{-19}$                                   & 2.53     \\
Al-Be         & $3.548\times10^{-19}$           & $3.555\times10^{-19}$                                   & 0.20       & Ir-V          & $4.301\times10^{-19}$               & $4.346\times10^{-19}$                                   & 1.05     \\
Al-Cu         & $3.388\times10^{-19}$           & $3.467\times10^{-19}$                                   & 2.33       & Ir-W          & $5.221\times10^{-19}$               & $5.223\times10^{-19}$                                   & 0.04     \\
Al-Ir         & $4.166\times10^{-19}$           & $4.357\times10^{-19}$                                   & 4.58       & Ir-Zr         & $4.009\times10^{-19}$               & $4.065\times10^{-19}$                                   & 1.40     \\
Al-Mo         & $4.021\times10^{-19}$           & $4.133\times10^{-19}$                                   & 2.79       & Mo-Ni         & $4.195\times10^{-19}$               & $4.212\times10^{-19}$                                   & 0.41     \\
Al-Nb         & $3.933\times10^{-19}$           & $4.060\times10^{-19}$                                   & 3.23       & Mo-Ta         & $4.642\times10^{-19}$               & $4.648\times10^{-19}$                                   & 0.13     \\
Al-Ti         & $3.041\times10^{-19}$           & $3.079\times10^{-19}$                                   & 1.25       & Mo-Ti         & $3.525\times10^{-19}$               & $3.581\times10^{-19}$                                   & 1.59     \\
Al-W          & $4.074\times10^{-19}$           & $4.260\times10^{-19}$                                   & 4.57       & Mo-V          & $4.105\times10^{-19}$               & $4.122\times10^{-19}$                                   & 0.41     \\
Au-Cu         & $3.676\times10^{-19}$           & $3.686\times10^{-19}$                                   & 0.27       & Mo-W          & $4.942\times10^{-19}$               & $4.954\times10^{-19}$                                   & 0.24     \\
Au-Hf         & $3.314\times10^{-19}$           & $3.327\times10^{-19}$                                   & 0.39       & Mo-Zr         & $3.828\times10^{-19}$               & $3.856\times10^{-19}$                                   & 0.73     \\
Au-Ir         & $4.620\times10^{-19}$           & $4.633\times10^{-19}$                                   & 0.28       & Nb-Ti         & $3.463\times10^{-19}$               & $3.518\times10^{-19}$                                   & 1.59     \\
Au-Mo         & $4.377\times10^{-19}$           & $4.394\times10^{-19}$                                   & 0.39       & Nb-W          & $4.859\times10^{-19}$               & $4.866\times10^{-19}$                                   & 0.14     \\
Au-Rh         & $4.275\times10^{-19}$           & $4.280\times10^{-19}$                                   & 0.12       & Nb-Zr         & $3.761\times10^{-19}$               & $3.788\times10^{-19}$                                   & 0.72     \\
Au-Ti         & $3.229\times10^{-19}$           & $3.274\times10^{-19}$                                   & 1.39       & Ni-Pt         & $4.067\times10^{-19}$               & $4.076\times10^{-19}$                                   & 0.22     \\
Au-W          & $4.518\times10^{-19}$           & $4.529\times10^{-19}$                                   & 0.24       & Ni-Rh         & $4.095\times10^{-19}$               & $4.103\times10^{-19}$                                   & 0.20     \\
Be-Cu         & $3.416\times10^{-19}$           & $3.468\times10^{-19}$                                   & 1.52       & Ni-Ta         & $4.067\times10^{-19}$               & $4.074\times10^{-19}$                                   & 0.17     \\
Be-Fe         & $3.671\times10^{-19}$           & $3.716\times10^{-19}$                                   & 1.23       & Ni-Ti         & $3.115\times10^{-19}$               & $3.139\times10^{-19}$                                   & 0.77     \\
Be-Ir         & $4.223\times10^{-19}$           & $4.358\times10^{-19}$                                   & 3.20       & Ni-V          & $3.606\times10^{-19}$               & $3.613\times10^{-19}$                                   & 0.19     \\
Be-Mo         & $4.066\times10^{-19}$           & $4.134\times10^{-19}$                                   & 1.67       & Ni-W          & $4.318\times10^{-19}$               & $4.342\times10^{-19}$                                   & 0.56     \\
Be-Ni         & $3.560\times10^{-19}$           & $3.623\times10^{-19}$                                   & 1.77       & Os-Pt         & $4.643\times10^{-19}$               & $4.646\times10^{-19}$                                   & 0.06     \\
Be-Nb         & $3.981\times10^{-19}$           & $4.061\times10^{-19}$                                   & 2.01       & Os-Ta         & $4.637\times10^{-19}$               & $4.643\times10^{-19}$                                   & 0.13     \\
Be-Ta         & $3.913\times10^{-19}$           & $3.998\times10^{-19}$                                   & 2.17       & Os-Ti         & $3.494\times10^{-19}$               & $3.577\times10^{-19}$                                   & 2.38     \\
Be-Ti         & $3.055\times10^{-19}$           & $3.080\times10^{-19}$                                   & 0.82       & Os-V          & $4.077\times10^{-19}$               & $4.118\times10^{-19}$                                   & 1.01     \\
Be-W          & $4.132\times10^{-19}$           & $4.261\times10^{-19}$                                   & 3.12       & Os-W          & $4.948\times10^{-19}$               & $4.949\times10^{-19}$                                   & 0.02     \\
Cr-Ir         & $4.401\times10^{-19}$           & $4.451\times10^{-19}$                                   & 1.14       & Os-Zr         & $3.803\times10^{-19}$               & $3.852\times10^{-19}$                                   & 1.29     \\
Cr-Mo         & $4.203\times10^{-19}$           & $4.222\times10^{-19}$                                   & 0.45       & Pt-Ti         & $3.404\times10^{-19}$               & $3.465\times10^{-19}$                                   & 1.79     \\
Cr-Pt         & $4.054\times10^{-19}$           & $4.086\times10^{-19}$                                   & 0.79       & Pt-W          & $4.789\times10^{-19}$               & $4.794\times10^{-19}$                                   & 0.10     \\
Cr-Ti         & $3.133\times10^{-19}$           & $3.146\times10^{-19}$                                   & 0.41       & Pt-Zr         & $3.698\times10^{-19}$               & $3.732\times10^{-19}$                                   & 0.92     \\
Cr-W          & $4.305\times10^{-19}$           & $4.352\times10^{-19}$                                   & 1.09       & Rh-Ta         & $4.526\times10^{-19}$               & $4.527\times10^{-19}$                                   & 0.02     \\
Cu-Ir         & $4.211\times10^{-19}$           & $4.250\times10^{-19}$                                   & 0.93       & Rh-Ti         & $3.438\times10^{-19}$               & $3.488\times10^{-19}$                                   & 1.45     \\
Cu-Mo         & $4.007\times10^{-19}$           & $4.032\times10^{-19}$                                   & 0.62       & Rh-W          & $4.817\times10^{-19}$               & $4.825\times10^{-19}$                                   & 0.17     \\
Cu-Pt         & $3.883\times10^{-19}$           & $3.902\times10^{-19}$                                   & 0.49       & Ta-Ti         & $3.412\times10^{-19}$               & $3.463\times10^{-19}$                                   & 1.49     \\
Cu-Ti         & $2.990\times10^{-19}$           & $3.004\times10^{-19}$                                   & 0.47       & Ta-W          & $4.784\times10^{-19}$               & $4.791\times10^{-19}$                                   & 0.15     \\
Cu-W          & $4.118\times10^{-19}$           & $4.156\times10^{-19}$                                   & 0.92       & Ta-Zr         & $3.704\times10^{-19}$               & $3.729\times10^{-19}$                                   & 0.67     \\
Fe-Hf         & $3.259\times10^{-19}$           & $3.271\times10^{-19}$                                   & 0.37       & Ti-W          & $3.602\times10^{-19}$               & $3.691\times10^{-19}$                                   & 2.47     \\
Fe-Ir         & $4.528\times10^{-19}$           & $4.554\times10^{-19}$                                   & 0.57       & Ti-Zr         & $2.868\times10^{-19}$               & $2.873\times10^{-19}$                                   & 0.17     \\
Fe-Mo         & $4.311\times10^{-19}$           & $4.320\times10^{-19}$                                   & 0.21       & V-W           & $4.207\times10^{-19}$               & $4.249\times10^{-19}$                                   & 1.00     \\
Fe-Nb         & $4.235\times10^{-19}$           & $4.244\times10^{-19}$                                   & 0.21       & W-Zr          & $3.922\times10^{-19}$               & $3.975\times10^{-19}$                                   & 1.35     \\
\hline\hline
\end{tabular}
\end{table*}

\begin{table}[!t]
\centering
\caption{Dominant dielectric contribution to non-retarded Hamaker constants between $26$ identical elemental polycrystalline metals that are embedded in pure water. Exact Lifshitz theory results obtained with the full spectral method for metals using a Drude low frequency extrapolation and with the simple spectral method for water using three different dielectric representations; namely the Parsegian-Weiss ($A_{\mathrm{PW}}^{\mathrm{d}}$), Roth-Lenhoff ($A_{\mathrm{RL}}^{\mathrm{d}}$), Fiedler \emph{et al.} ($A_{\mathrm{FI}}^{\mathrm{d}}$) parameterizations.}\label{hamakerrecommended_water}
\begin{tabular}{c c c c}
\\ \hline\hline
Metal       & $A_{\mathrm{PW}}^{\mathrm{d}}$   & $A_{\mathrm{RL}}^{\mathrm{d}}$     & $A_{\mathrm{FI}}^{\mathrm{d}}$     \\
            & (Joule)                          & (Joule)                            & (Joule)                            \\ \hline\hline
Ag			& $2.394\times10^{-19}$            & $2.209\times10^{-19}$    	        & $2.119\times10^{-19}$              \\
Al			& $2.568\times10^{-19}$            & $2.436\times10^{-19}$    	        & $2.413\times10^{-19}$              \\
Au			& $2.679\times10^{-19}$            & $2.481\times10^{-19}$    	        & $2.386\times10^{-19}$              \\
Ba			& $1.003\times10^{-19}$            & $0.923\times10^{-19}$    	        & $0.937\times10^{-19}$              \\
Be			& $2.455\times10^{-19}$            & $2.303\times10^{-19}$    	        & $2.272\times10^{-19}$              \\
Co			& $2.780\times10^{-19}$            & $2.584\times10^{-19}$    	        & $2.496\times10^{-19}$              \\
Cr			& $2.500\times10^{-19}$            & $2.328\times10^{-19}$    	        & $2.266\times10^{-19}$              \\
Cu			& $2.154\times10^{-19}$            & $1.982\times10^{-19}$    	        & $1.915\times10^{-19}$              \\
Fe			& $2.610\times10^{-19}$            & $2.425\times10^{-19}$    	        & $2.350\times10^{-19}$              \\
Hf			& $1.553\times10^{-19}$            & $1.393\times10^{-19}$    	        & $1.340\times10^{-19}$              \\
Ir			& $3.975\times10^{-19}$            & $3.759\times10^{-19}$    	        & $3.632\times10^{-19}$              \\
Mo			& $3.501\times10^{-19}$            & $3.301\times10^{-19}$    	        & $3.203\times10^{-19}$              \\
Nb			& $3.296\times10^{-19}$            & $3.091\times10^{-19}$    	        & $2.992\times10^{-19}$              \\
Ni			& $2.416\times10^{-19}$            & $2.233\times10^{-19}$    	        & $2.155\times10^{-19}$              \\
Os			& $3.372\times10^{-19}$            & $3.152\times10^{-19}$    	        & $3.038\times10^{-19}$              \\
Pd			& $2.549\times10^{-19}$            & $2.354\times10^{-19}$    	        & $2.264\times10^{-19}$              \\
Pt			& $3.132\times10^{-19}$            & $2.925\times10^{-19}$    	        & $2.817\times10^{-19}$              \\
Rh			& $3.242\times10^{-19}$            & $3.042\times10^{-19}$    	        & $2.941\times10^{-19}$              \\
Sc			& $1.336\times10^{-19}$            & $1.214\times10^{-19}$    	        & $1.205\times10^{-19}$              \\
Sr			& $1.243\times10^{-19}$            & $1.132\times10^{-19}$    	        & $1.128\times10^{-19}$              \\
Ta			& $3.160\times10^{-19}$            & $2.958\times10^{-19}$    	        & $2.858\times10^{-19}$              \\
Ti			& $1.566\times10^{-19}$            & $1.424\times10^{-19}$    	        & $1.391\times10^{-19}$              \\
Tm			& $2.141\times10^{-19}$            & $1.980\times10^{-19}$    	        & $1.935\times10^{-19}$              \\
V			& $2.305\times10^{-19}$            & $2.131\times10^{-19}$    	        & $2.072\times10^{-19}$              \\
W			& $3.705\times10^{-19}$            & $3.486\times10^{-19}$    	        & $3.367\times10^{-19}$		         \\   	
Zr			& $1.896\times10^{-19}$            & $1.733\times10^{-19}$		        & $1.682\times10^{-19}$              \\ \hline\hline
\end{tabular}
\end{table}

\subsection{Identical metals in water}\label{water}

\noindent Measurements of the van der Waals forces are often carried out in water in order to eliminate capillary forces\,\cite{AFMrefe1,AFMrefe3}. Such forces can be dominant under ambient conditions and arise due to the thin layers of water vapor that are omnipresent in surfaces. This motivated full Lifshitz calculations of non-retarded Hamaker constants between identical metals embedded in room temperature water.

\begin{figure}
\centering
\includegraphics[width = 3.2in]{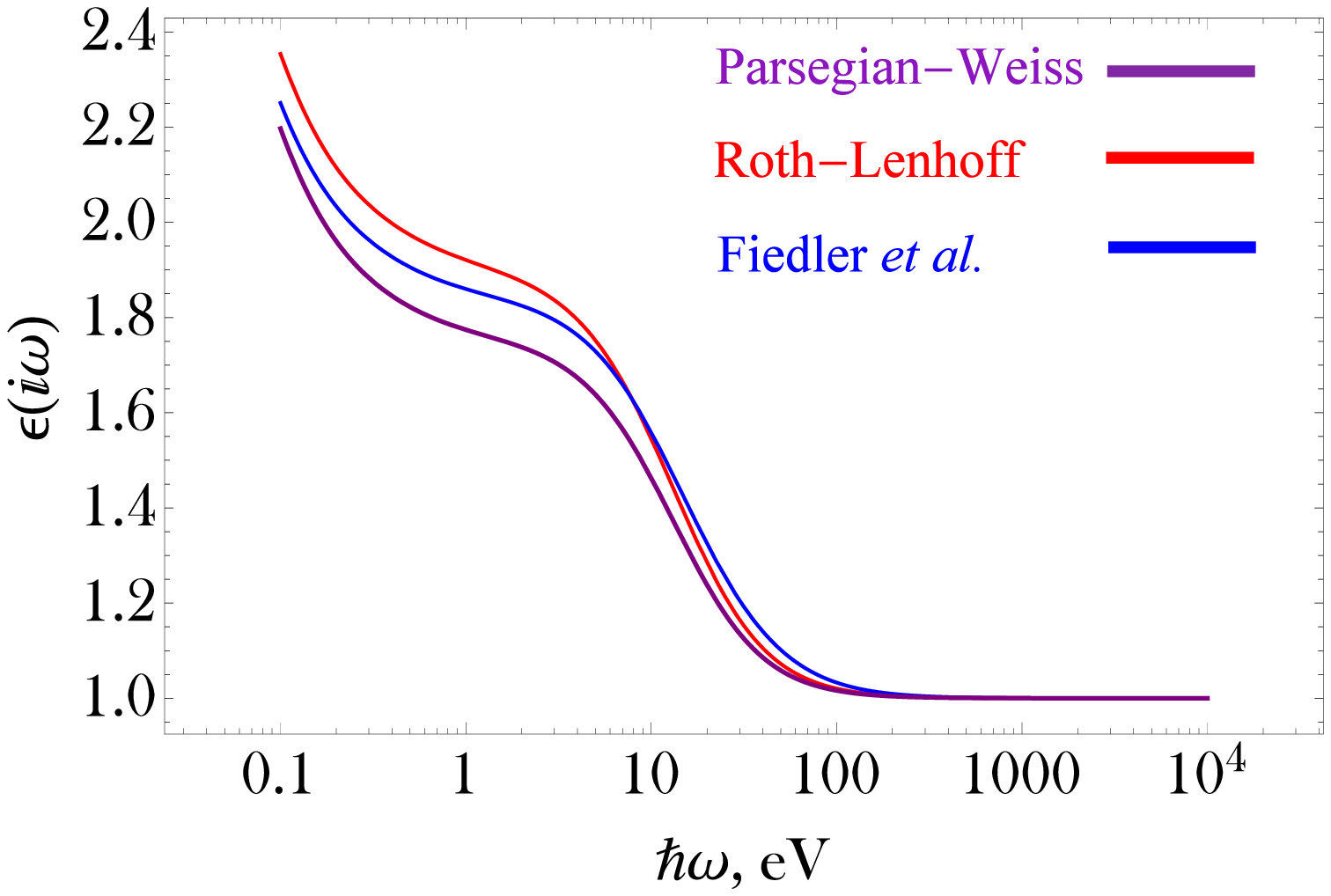}
\caption{Imaginary argument dielectric function of pure room temperature water according to three dielectric representations. Logarithmic plot in a photon energy range that contains all room temperature Matsubara frequencies considered in the Lifshitz calculations.}\label{waterfigure}
\end{figure}

The simple spectral method will be utilized for the dielectric function of pure water at room temperature. Three different parameterizations will be investigated. \textbf{(i)} The classic Parsegian-Weiss representation that is based on experimental data within the $0-25\,$eV interval and employs one Debye term in the microwave, five Lorenz terms in the infra-red and six Lorenz terms in the ultraviolet range\,\cite{waterdi1}. The fitting parameters are adopted from Parsegian's handbook\,\cite{Metalli2} and are slightly updated compared to the original values. \textbf{(ii)} The standard Roth-Lenhoff representation that is based on the same experimental $0-25\,$eV data and employs the same Debye term in the microwave, the same five Lorenz terms in the infra-red and six updated Lorenz terms in the ultraviolet range\,\cite{waterdi2}. \textbf{(iii)} The modern representation of Fiedler \emph{et al.}\,\cite{waterdi3} that is based on experimental data in the $0-100\,$eV interval benefitting from state-of-the-art VUV measurements using small-angle inelastic X-ray scattering\,\cite{waterdi4,waterdi5}. This extended representation consists of two Debye terms in the microwave, seven Lorenz terms in the infra-red and twelve Lorenz terms in the ultraviolet. The differences between these three representations are depicted in figure \ref{waterfigure}, where the imaginary argument dielectric functions have been plotted as functions of the photon energy. The Parsegian-Weiss result is the lowest in the whole range, whereas the Roth-Lenhoff result is higher than the Fiedler result at the longer wavelengths and lower than the Fiedler result at the intermediate and short wavelengths. The same can be deduced from the non-retarded Hamaker constants of the water-vacuum-water system, whose exact values are $3.890\times10^{-20}\,$J (P-W), $4.998\times10^{-20}\,$J (R-L), $5.378\times10^{-20}\,$J (F).

Exact Lifshitz theory results for non-retarded Hamaker constants between $26$ identical elemental isotropic metals that are immersed in water are given in Table \ref{hamakerrecommended_water} for the three dielectric representations of pure room temperature water. With the exception of the Ba alkaline earth metal that possesses the smallest $\epsilon(\imath\omega)$ amongst the investigated elements, we have $A_{\mathrm{PW}}^{\mathrm{d}}>A_{\mathrm{RL}}^{\mathrm{d}}>A_{\mathrm{FI}}^{\mathrm{d}}$. In particular, the mean, maximum and minimum absolute relative deviations between the Parsegian-Weiss and Fiedler representations are $10.9\%$, $15.9\%$, $6.4\%$ due to $\epsilon_{\mathrm{PW}}(\imath\omega)<\epsilon_{\mathrm{Fi}}(\imath\omega)$ being valid at all frequencies. On the other hand, the mean, maximum and minimum absolute relative deviations between the Roth-Lenhoff and Fiedler representations are merely $2.9\%$, $4.2\%$ and $0.4\%$ due to the $\sim10\,$eV crossover between the $\epsilon(\imath\omega)$ curves. The non-retarded Hamaker constants resulting from the Fiedler representation will be recommended for use, since it is based on the most extended-in-frequency dielectric dataset of water available.

\begin{table*}
\centering
\caption{Dominant dielectric contribution to the non-retarded Hamaker constants of $100$ elemental polycrystalline metal combinations embedded in water. Exact Lifshitz theory results obtained with the full spectral method for metals using a Drude low frequency extrapolation and the simple spectral method for water using the Fiedler \emph{et al.} dielectric representation. These first principle results are compared with the results of the standard combining relation, see Eq.\ref{combiningrule2}. Note that reciprocity applies as also dictated from Newton's third law, $A_{1\mathrm{w}2}=A_{2\mathrm{w}1}$.}\label{hamakerrecommended_combinationFI}
\begin{tabular}{c c c c || c c c c}
\\ \hline\hline
Pair   & $A_{1\mathrm{w}2}^{\mathrm{d}}$ (Joule) & $\sqrt{A_{1\mathrm{w}1}^{\mathrm{d}}A_{2\mathrm{w}2}^{\mathrm{d}}}$ (Joule) & Dev $(\%)$ & Pair   & $A_{1\mathrm{w}2}^{\mathrm{d}}$ (Joule) & $\sqrt{A_{1\mathrm{w}1}^{\mathrm{d}}A_{2\mathrm{w}2}^{\mathrm{d}}}$ (Joule) & Dev $(\%)$    \\ \hline\hline
Ag-Al  &        $2.101\times10^{-19}$            &     $2.261\times10^{-19}$                                                   &  $7.62$    & Fe-Ni  &        $2.246\times10^{-19}$            &     $2.250\times10^{-19}$                                                   &     $0.18$    \\
Ag-Au  &        $2.243\times10^{-19}$            &     $2.249\times10^{-19}$                                                   &  $0.27$    & Fe-Ta  &        $2.582\times10^{-19}$            &     $2.592\times10^{-19}$                                                   &     $0.39$    \\
Ag-Co  &        $2.284\times10^{-19}$            &     $2.300\times10^{-19}$                                                   &  $0.70$    & Fe-V   &        $2.203\times10^{-19}$            &     $2.207\times10^{-19}$                                                   &     $0.18$    \\
Ag-Cu  &        $2.005\times10^{-19}$            &     $2.014\times10^{-19}$                                                   &  $0.45$    & Fe-W   &        $2.777\times10^{-19}$            &     $2.813\times10^{-19}$                                                   &     $1.30$    \\
Ag-Fe  &        $2.213\times10^{-19}$            &     $2.232\times10^{-19}$                                                   &  $0.86$    & Hf-Ir  &        $2.128\times10^{-19}$            &     $2.206\times10^{-19}$                                                   &     $3.67$    \\
Ag-Ir  &        $2.728\times10^{-19}$            &     $2.774\times10^{-19}$                                                   &  $1.69$    & Hf-Mo  &        $2.011\times10^{-19}$            &     $2.072\times10^{-19}$                                                   &     $3.03$    \\
Ag-Mo  &        $2.558\times10^{-19}$            &     $2.605\times10^{-19}$                                                   &  $1.84$    & Hf-Nb  &        $1.951\times10^{-19}$            &     $2.002\times10^{-19}$                                                   &     $2.61$    \\
Ag-Ni  &        $2.130\times10^{-19}$            &     $2.137\times10^{-19}$                                                   &  $0.33$    & Hf-Os  &        $1.953\times10^{-19}$            &     $2.018\times10^{-19}$                                                   &     $3.33$    \\
Ag-Pd  &        $2.185\times10^{-19}$            &     $2.190\times10^{-19}$                                                   &  $0.23$    & Hf-Ta  &        $1.918\times10^{-19}$            &     $1.957\times10^{-19}$                                                   &     $2.03$    \\
Ag-Pt  &        $2.426\times10^{-19}$            &     $2.443\times10^{-19}$                                                   &  $0.70$    & Hf-W   &        $2.049\times10^{-19}$            &     $2.124\times10^{-19}$                                                   &     $3.66$    \\
Ag-Ta  &        $2.439\times10^{-19}$            &     $2.461\times10^{-19}$                                                   &  $0.90$    & Ir-Pd  &        $2.844\times10^{-19}$            &     $2.868\times10^{-19}$                                                   &     $0.84$    \\
Ag-Ti  &        $1.668\times10^{-19}$            &     $1.717\times10^{-19}$                                                   &  $2.94$    & Ir-Rh  &        $3.257\times10^{-19}$            &     $3.268\times10^{-19}$                                                   &     $0.34$    \\
Ag-W   &        $2.625\times10^{-19}$            &     $2.671\times10^{-19}$                                                   &  $1.75$    & Ir-Ti  &        $2.101\times10^{-19}$            &     $2.248\times10^{-19}$                                                   &     $7.00$    \\
Al-Be  &        $2.332\times10^{-19}$            &     $2.341\times10^{-19}$                                                   &  $0.39$    & Ir-V   &        $2.681\times10^{-19}$            &     $2.743\times10^{-19}$                                                   &     $2.31$    \\
Al-Cu  &        $2.060\times10^{-19}$            &     $2.150\times10^{-19}$                                                   &  $4.37$    & Ir-W   &        $3.495\times10^{-19}$            &     $3.497\times10^{-19}$                                                   &     $0.06$    \\
Al-Ir  &        $2.715\times10^{-19}$            &     $2.960\times10^{-19}$                                                   &  $9.02$    & Ir-Zr  &        $2.387\times10^{-19}$            &     $2.472\times10^{-19}$                                                   &     $3.56$    \\
Al-Mo  &        $2.632\times10^{-19}$            &     $2.780\times10^{-19}$                                                   &  $5.62$    & Mo-Ni  &        $2.600\times10^{-19}$            &     $2.627\times10^{-19}$                                                   &     $1.04$    \\
Al-Nb  &        $2.524\times10^{-19}$            &     $2.687\times10^{-19}$                                                   &  $6.46$    & Mo-Ta  &        $3.018\times10^{-19}$            &     $3.026\times10^{-19}$                                                   &     $0.27$    \\
Al-Ti  &        $1.786\times10^{-19}$            &     $1.832\times10^{-19}$                                                   &  $2.58$    & Mo-Ti  &        $2.013\times10^{-19}$            &     $2.111\times10^{-19}$                                                   &     $4.87$    \\
Al-W   &        $2.609\times10^{-19}$            &     $2.850\times10^{-19}$                                                   &  $9.24$    & Mo-V   &        $2.548\times10^{-19}$            &     $2.576\times10^{-19}$                                                   &     $1.10$    \\
Au-Cu  &        $2.123\times10^{-19}$            &     $2.138\times10^{-19}$                                                   &  $0.71$    & Mo-W   &        $3.269\times10^{-19}$            &     $3.284\times10^{-19}$                                                   &     $0.46$    \\
Au-Hf  &        $1.766\times10^{-19}$            &     $1.788\times10^{-19}$                                                   &  $1.25$    & Mo-Zr  &        $2.272\times10^{-19}$            &     $2.321\times10^{-19}$                                                   &     $2.16$    \\
Au-Ir  &        $2.922\times10^{-19}$            &     $2.944\times10^{-19}$                                                   &  $0.75$    & Nb-Ti  &        $1.947\times10^{-19}$            &     $2.040\times10^{-19}$                                                   &     $4.78$    \\
Au-Mo  &        $2.740\times10^{-19}$            &     $2.764\times10^{-19}$                                                   &  $0.88$    & Nb-W   &        $3.165\times10^{-19}$            &     $3.174\times10^{-19}$                                                   &     $0.28$    \\
Au-Rh  &        $2.641\times10^{-19}$            &     $2.649\times10^{-19}$                                                   &  $0.30$    & Nb-Zr  &        $2.198\times10^{-19}$            &     $2.243\times10^{-19}$                                                   &     $2.05$    \\
Au-Ti  &        $1.756\times10^{-19}$            &     $1.822\times10^{-19}$                                                   &  $3.76$    & Ni-Pt  &        $2.450\times10^{-19}$            &     $2.464\times10^{-19}$                                                   &     $0.57$    \\
Au-W   &        $2.813\times10^{-19}$            &     $2.834\times10^{-19}$                                                   &  $0.75$    & Ni-Rh  &        $2.506\times10^{-19}$            &     $2.518\times10^{-19}$                                                   &     $0.48$    \\
Be-Cu  &        $2.029\times10^{-19}$            &     $2.086\times10^{-19}$                                                   &  $2.81$    & Ni-Ta  &        $2.470\times10^{-19}$            &     $2.482\times10^{-19}$                                                   &     $0.49$    \\
Be-Fe  &        $2.259\times10^{-19}$            &     $2.311\times10^{-19}$                                                   &  $2.30$    & Ni-Ti  &        $1.694\times10^{-19}$            &     $1.731\times10^{-19}$                                                   &     $2.18$    \\
Be-Ir  &        $2.704\times10^{-19}$            &     $2.873\times10^{-19}$                                                   &  $6.25$    & Ni-V   &        $2.104\times10^{-19}$            &     $2.113\times10^{-19}$                                                   &     $0.43$    \\
Be-Mo  &        $2.609\times10^{-19}$            &     $2.698\times10^{-19}$                                                   &  $3.41$    & Ni-W   &        $2.655\times10^{-19}$            &     $2.694\times10^{-19}$                                                   &     $1.47$    \\
Be-Ni  &        $2.142\times10^{-19}$            &     $2.213\times10^{-19}$                                                   &  $3.31$    & Os-Pt  &        $2.920\times10^{-19}$            &     $2.925\times10^{-19}$                                                   &     $0.17$    \\
Be-Nb  &        $2.507\times10^{-19}$            &     $2.607\times10^{-19}$                                                   &  $3.99$    & Os-Ta  &        $2.937\times10^{-19}$            &     $2.947\times10^{-19}$                                                   &     $0.34$    \\
Be-Ta  &        $2.447\times10^{-19}$            &     $2.548\times10^{-19}$                                                   &  $4.13$    & Os-Ti  &        $1.925\times10^{-19}$            &     $2.056\times10^{-19}$                                                   &     $6.81$    \\
Be-Ti  &        $1.744\times10^{-19}$            &     $1.778\times10^{-19}$                                                   &  $1.95$    & Os-V   &        $2.452\times10^{-19}$            &     $2.509\times10^{-19}$                                                   &     $2.32$    \\
Be-W   &        $2.602\times10^{-19}$            &     $2.766\times10^{-19}$                                                   &  $6.30$    & Os-W   &        $3.197\times10^{-19}$            &     $3.198\times10^{-19}$                                                   &     $0.03$    \\
Cr-Ir  &        $2.801\times10^{-19}$            &     $2.869\times10^{-19}$                                                   &  $2.43$    & Os-Zr  &        $2.186\times10^{-19}$            &     $2.261\times10^{-19}$                                                   &     $3.43$    \\
Cr-Mo  &        $2.665\times10^{-19}$            &     $2.694\times10^{-19}$                                                   &  $1.09$    & Pt-Ti  &        $1.886\times10^{-19}$            &     $1.980\times10^{-19}$                                                   &     $4.98$    \\
Cr-Pt  &        $2.486\times10^{-19}$            &     $2.527\times10^{-19}$                                                   &  $1.65$    & Pt-W   &        $3.072\times10^{-19}$            &     $3.080\times10^{-19}$                                                   &     $0.26$    \\
Cr-Ti  &        $1.750\times10^{-19}$            &     $1.775\times10^{-19}$                                                   &  $1.43$    & Pt-Zr  &        $2.129\times10^{-19}$            &     $2.177\times10^{-19}$                                                   &     $2.25$    \\
Cr-W   &        $2.696\times10^{-19}$            &     $2.762\times10^{-19}$                                                   &  $2.45$    & Rh-Ta  &        $2.897\times10^{-19}$            &     $2.899\times10^{-19}$                                                   &     $0.07$    \\
Cu-Ir  &        $2.575\times10^{-19}$            &     $2.637\times10^{-19}$                                                   &  $2.41$    & Rh-Ti  &        $1.941\times10^{-19}$            &     $2.023\times10^{-19}$                                                   &     $4.22$    \\
Cu-Mo  &        $2.435\times10^{-19}$            &     $2.477\times10^{-19}$                                                   &  $1.72$    & Rh-W   &        $3.134\times10^{-19}$            &     $3.147\times10^{-19}$                                                   &     $0.41$    \\
Cu-Pt  &        $2.293\times10^{-19}$            &     $2.323\times10^{-19}$                                                   &  $1.31$    & Ta-Ti  &        $1.911\times10^{-19}$            &     $1.994\times10^{-19}$                                                   &     $4.34$    \\
Cu-Ti  &        $1.612\times10^{-19}$            &     $1.632\times10^{-19}$                                                   &  $1.24$    & Ta-W   &        $3.092\times10^{-19}$            &     $3.102\times10^{-19}$                                                   &     $0.32$    \\
Cu-W   &        $2.478\times10^{-19}$            &     $2.539\times10^{-19}$                                                   &  $2.46$    & Ta-Zr  &        $2.153\times10^{-19}$            &     $2.193\times10^{-19}$                                                   &     $1.86$    \\
Fe-Hf  &        $1.755\times10^{-19}$            &     $1.775\times10^{-19}$                                                   &  $1.14$    & Ti-W   &        $2.023\times10^{-19}$            &     $2.164\times10^{-19}$                                                   &     $6.97$    \\
Fe-Ir  &        $2.885\times10^{-19}$            &     $2.922\times10^{-19}$                                                   &  $1.28$    & Ti-Zr  &        $1.521\times10^{-19}$            &     $1.530\times10^{-19}$                                                   &     $0.59$    \\
Fe-Mo  &        $2.729\times10^{-19}$            &     $2.744\times10^{-19}$                                                   &  $0.55$    & V-W    &        $2.581\times10^{-19}$            &     $2.641\times10^{-19}$                                                   &     $2.32$    \\
Fe-Nb  &        $2.639\times10^{-19}$            &     $2.652\times10^{-19}$                                                   &  $0.49$    & W-Zr   &        $2.299\times10^{-19}$            &     $2.380\times10^{-19}$                                                   &     $3.52$    \\
\hline\hline
\end{tabular}
\end{table*}

\begin{figure*}
\centering
\includegraphics[width = 7.0in]{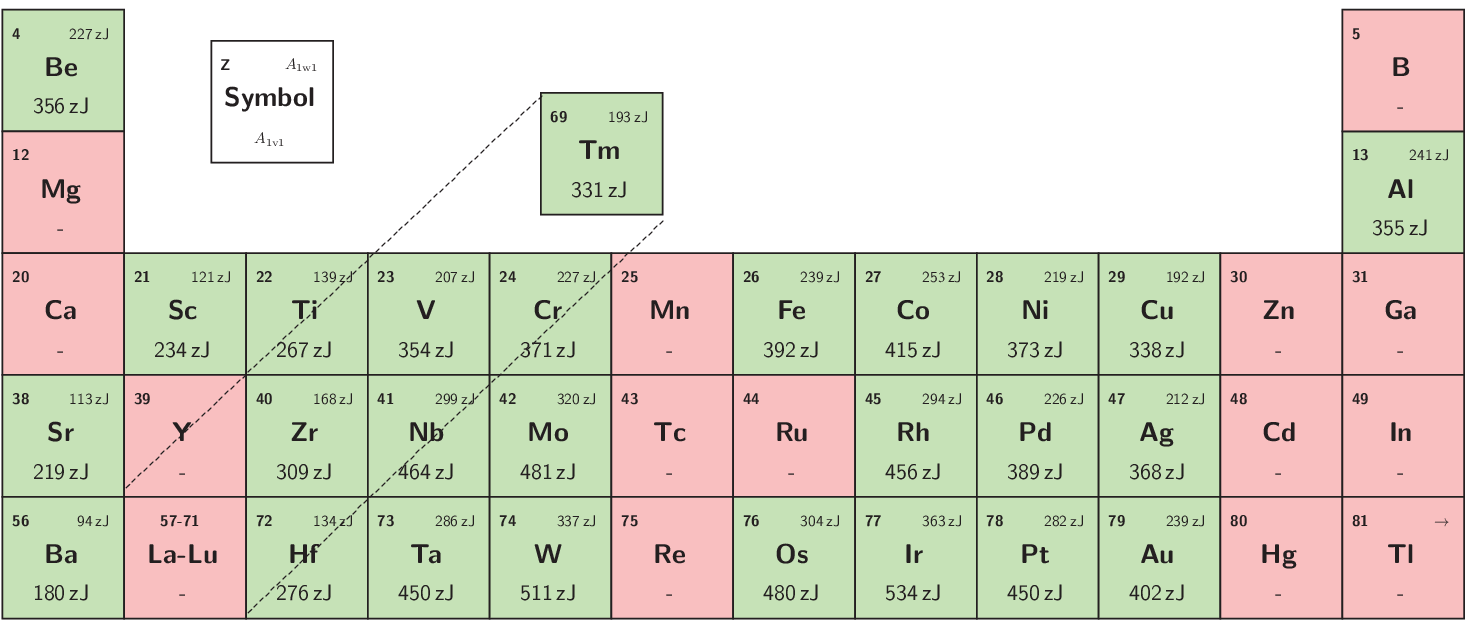}
\caption{The periodic table of non-retarded room temperature Hamaker constants between $26$ identical isotropic polycrystalline metals that are embedded in vacuum and pure water. Exact Lifshitz theory results that consider both dielectric and magnetic contributions (the latter are non-negligible only for ferromagnetic materials). The imaginary argument dielectric function of the metals has been obtained with the full spectral method; a Drude low frequency extrapolation has been employed and the necessary extended-in-frequency dielectric data ranging up to $10000$eV have been adopted from the handbook of Adachi\,\cite{opticalB}. The imaginary argument dielectric function of pure water has been obtained with the simple spectral method; the Fiedler \emph{et al.} representation has been employed that is based on experimental data up to $100\,$eV\,\cite{waterdi3}. The element boxes contain the atomic number (left), the recommended Hamaker constant through water in zJ (right) and the recommended Hamaker constant through vacuum in zJ (bottom).}\label{periodictable}
\end{figure*}

\subsection{Metal combinations in water}\label{differentwater}

\noindent In the presence of a surrounding medium such as water, the geometric combining relation can be straightforwardly generalized to\,\cite{Metalli1}
\begin{equation}
A_{1\mathrm{w}2}^{\mathrm{d}}\simeq\sqrt{A_{1\mathrm{w}1}^{\mathrm{d}}A_{2\mathrm{w}2}^{\mathrm{d}}}\,.\label{combiningrule2}
\end{equation}
The accuracy of the geometric mixing rule has been tested against full Lifshitz computations for $100$ metal combinations embedded in water. The geometric mean is revealed to be an accurate approximation for all aforementioned dielectric representations of water. For the Fiedler \emph{et al.} parameterization, see  Table \ref{hamakerrecommended_combinationFI} for details, the mean, maximum and minimum relative deviations with respect to the exact result are $2.25\%$, $9.24\%$ and $0.03\%$, respectively. For the Parsegian-Weiss parameterization, the mean, maximum and minimum relative deviations with respect to the exact result are $1.81\%$, $7.87\%$ and $0.05\%$, respectively. For the Roth-Lenhoff parameterization, the mean, maximum and minimum relative deviations with respect to the exact result are  $2.05\%$, $8.65\%$ and $0.06\%$, respectively. The geometric mixing rule retains its high accuracy, but is somewhat less precise in pure water compared to vacuum.

\section{Summary and discussion}\label{outro}

\noindent Exact Lifshitz calculations have been reported for the non-retarded room temperature Hamaker constants between $26$ identical isotropic polycrystalline metals embedded in vacuum and pure water. \emph{For metals}, the full spectral method complemented with Drude low frequency extrapolation has been employed with input from state-of-the-art extended-in-frequency dielectric data that range from the far infra-red up to the soft X-ray region of the electromagnetic spectrum, \emph{i.e.} $\hbar\omega=10\,$keV. The upper numerical cut-offs imposed on the infinite summations and the Kramers-Kronig integrals were verified to lead to negligible errors, whereas the results were demonstrated to be weakly sensitive to the low frequency extrapolation procedure. Magnetic contributions were also considered, but proved to be either small for ferromagnetic metals or negligible for paramagnetic or diamagnetic materials. \emph{For pure water}, the simple spectral method has been employed adopting the recent dielectric representation of Fiedler \emph{et al.} based on state-of-the-art experimental data up to $\hbar\omega=100\,$eV. Different dielectric parameterizations were also probed.

\emph{In the case of vacuum as the intervening medium}, non-retarded Hamaker constants are expected to be accurate within few percent, with the percentage depending on the material owing to the varying data quality and frequency extrapolation range. The low temperature approximation was demonstrated to be accurate within $1\%$ compared to the exact Lifshitz result. \emph{In the case of pure water as the intervening medium}, the non-retarded Hamaker constants are expected to be accurate well within $10\%$. The additional inaccuracies stem from fitting errors in the UV dominant part of the dielectric representation and the lack of pure water measurements beyond $100\,$eV. \emph{Overall}, the recommended values for the non-retarded room temperature Hamaker constants between the $26$ identical isotropic metals embedded in vacuum and in pure water have been gathered in figure \ref{periodictable}.  Finally, non-retarded Hamaker calculations have been reported for $100$ elemental isotropic metal combinations embedded in vacuum and water. The well-known geometric combining relation was demonstrated to be accurate within $1\%$ (vacuum) and $2\%$ (water) compared to the exact Lifshitz result.

To the author's knowledge, the present compilation features the most accurate non-retarded room temperature Hamaker constants of metals reported in the literature. As such, the recommended Hamaker constants can be compared with dedicated measurements. However, in the case of high precision measurements involving bodies covered with metallic films, it is always preferable to measure the magneto-dielectric response of the employed sample due to the non-negligible variations caused by microstructural differences\,\cite{outrore0}. In addition, the recommended values can be directly employed in the modelling of van der Waals interactions in different physical phenomena such as colloidal stability and powder adhesion. Moreover, the recommendations can serve as reference values in modern theoretical studies of implicit temperature effects in bulk metals\,\cite{outrore1,outrore2}, size effects in metal nano-particles\,\cite{outrore2,outrore3}, spatial dispersion effects in bulk metals\,\cite{Lifshit6,Lifshit8}, retardation effects at large distances\,\cite{abiniti1}, beyond step-like interface effects\,\cite{outrore4,outrore5} and inhomogeneity effects\,\cite{outrore6,outrore7,outrore8}. Our future work will focus on expanding the present non-retarded Hamaker compilation with the inclusion of additional elemental metals outside the Adachi database as well as on systematic calculations of the separation-dependent Hamaker coefficients of elemental metals based on retarded Lifshitz calculations.

\end{document}